%% file: main.tex
\documentclass[sigconf]{acmart}
\usepackage[utf8]{inputenc}

\usepackage{rotating}
\usepackage{adjustbox}
\usepackage{pdfpages}
\usepackage{caption}
\usepackage{subcaption}
\usepackage{soul}
\usepackage{array}
\usepackage{graphicx}
\graphicspath{ {./figures/} }
\usepackage{wrapfig}

\usepackage{color, xcolor,colortbl}
\usepackage{tabularx}
\usepackage{multirow}
\usepackage[utf8]{inputenc}
\usepackage{hyperref}

\AtBeginDocument{%
  \providecommand\BibTeX{{%
    \normalfont B\kern-0.5em{\scshape i\kern-0.25em b}\kern-0.8em\TeX}}}

\setcopyright{none}
\acmDOI{ 10.48550/arXiv.2508.08672}
\acmISBN{}

\acmConference[]{LIMITS '25}{11th Workshop on Computing Within Limits}{June 26-27, 2025, Online}
                  
\begin{document}

\title{Imposing AI: Deceptive design patterns against sustainability}

\author{Anaëlle Beignon}
\authornote{All authors contributed equally to this research.}
\email{a.beignon@unistra.fr}
\affiliation{%
  \institution{Université de Strasbourg}
  \city{Strasbourg}
  \country{France}
}

\author{Thomas Thibault}
\email{thomas@designersethiques.org}
\affiliation{%
  \institution{Designers Éthiques}
  \city{Paris}
  \country{France}
}

\author{Nolwenn Maudet}
\email{nmaudet@unistra.fr}
\affiliation{%
  \institution{Université de Strasbourg}
  \city{Strasbourg}
  \country{France}
}

\renewcommand{\shortauthors}{Beignon, Thibault \& Maudet}

\begin{abstract}
Generative AI is being massively deployed in digital services, at a scale that will result in significant environmental harm. We document how tech companies are transforming established user interfaces to impose AI use and show how and to what extent these strategies fit within established deceptive pattern categories. We identify two main design strategies that are implemented to impose AI use in both personal and professional contexts: imposing AI features in interfaces at the expense of existing non-AI features and promoting narratives about AI that make it harder to resist using it. We discuss opportunities for regulating the imposed adoption of AI features, which would inevitably lead to negative environmental effects.
\end{abstract}




\begin{teaserfigure} 
  \includegraphics[ width=\textwidth]{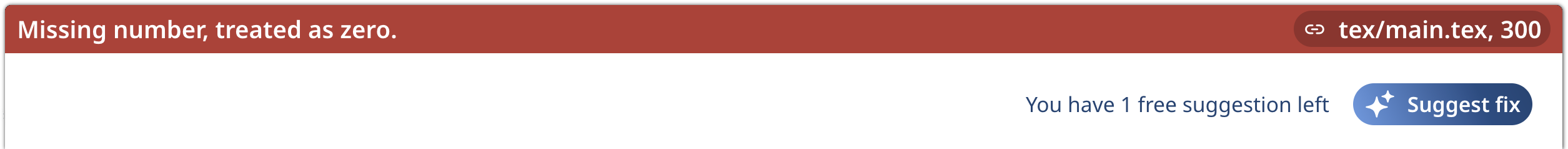}
  \caption{Overleaf offering us to fix this paper with AI.}
  \Description{A screenshot of the troubleshooting interface of Overleaf, which points at a line that contains an error. There is a button which says "Suggest fix" along with a spark Icon. A smaller text says "you have 1 free suggestion left".}
  \label{fig:teaser}
\end{teaserfigure}

\maketitle

\input{tex/1_introduction}

\input{tex/2_background}
\input{tex/3_method}
\input{tex/4_results}

\input{tex/5_discussion}
\input{tex/6_conclusion}

\bibliographystyle{ACM-Reference-Format}
\bibliography{main.bbl}

\end{document}

%% file: tex/1_introduction.tex
\section{Introduction}

\begin{quote}
\textit{“Computing is evolving again. We spoke last year about this important shift in computing from a mobile-first to an AI-first approach. … In an AI-first world, we are rethinking all our products and applying machine learning and AI to solve user problems.”}\newline Sundar Pichai keynote in 2017\footnote{https://www.youtube.com/live/Y2VF8tmLFHw?t=411s}
\end{quote}

We are currently witnessing a paradigm shift in the software industry, towards an AI-first mindset. Not only do many new companies develop AI-based products, but many existing tech companies are integrating AI-based features into their existing products. 

This massive turn towards generative AI has worrying social and environmental costs that researchers are just starting to estimate. Studying the environmental impact of an AI service based on Stable Diffusion, Berthelot et al. explain that \textit{“with 360 tons of carbon equivalent emission, an impact on metal scarcity equivalent to the production of 5659 smartphones, and an energy footprint of 2.48 Gigawatt hours, it is clear that the impact of Gen-AI should be a matter of concern and not only for its carbon footprint.”} \cite{berthelotEstimatingEnvironmentalImpact2024}. In addition to carbon emissions and mineral use, the massive development of AI requires large quantities of water, conflicting with human use and putting a strain on access to this key resource \cite{liMakingAILess2025}. Worse, AI is on a path to undermine all optimization efforts and instead worsen the environmental impact of ICT. During the first semester of 2023, North America set a record with a 25\% increase in the construction of new data centers\footnote{https://www.businessinsider.com/ai-data-centers-land-grab-google-meta-openai-amazon-2023-12} and a 2024 report by the International Energy Agency estimated that electricity usage by data centers \textit{"is set to more than double to around 945 TWh by 2030"}\cite{EnergyAIAnalysis2025}.

Following the cornucopian paradigm described by Preist et al.~\cite{preistUnderstandingMitigatingEffects2016}, tech companies are currently anticipating a growth in demand for their gen-AI products. Accordingly, they have started reshaping and building new infrastructure (data-centers, energy transport and production, etc.) to support this future expected demand, in the process worsening existing environmental issues. The concern for winning the "AI war" has become a national one: for example, the US plans to extend the lifespan of coal power plants to support the increasing energy demand due to AI \cite{chuUSSlowsPlans2024}. AI use has thus been confidently anticipated.

But will users actually adopt AI-based features?
This is not yet guaranteed.
Generative AI is beset by various controversies that could limit people’s willingness to adopt such features or pay for them \cite{baldassarreSocialImpactGenerative2023}. The fallibility of gen-AI has long been criticized, and is even debated in mainstream media outlets: a BBC study has for instance found that AI assistants give inaccurate responses more than half of the time \footnote{https://www.bbc.co.uk/aboutthebbc/documents/bbc-research-into-ai-assistants.pdf}. 
Moreover, many activist-led campaigns have sought to raise awareness about the social and environmental impacts of AI.\footnote{see for example https://savethe.ai/, https://www.bbc.com/news/articles/cwyd3r62kp5o } 
More broadly speaking, trust issues regarding AI drive \textit{"algorithm aversion"}\cite{casteloTaskDependentAlgorithmAversion2019} or \textit{"human favoritism"} \cite{zhangHumanFavoritismNot2023} and lower its appeal to potential users \cite{casteloTaskDependentAlgorithmAversion2019}.

To push for adoption, tech companies have been investing in large-scale marketing efforts, backed by extensive media coverage in which AI products and features are often presented as revolutionary. More surprisingly, in an arguably unprecedented development at this scale, companies have also been leveraging UX and UI design strategies to promote the adoption of AI-based features. In early 2024, multiple AI buttons began appearing into our daily digital tools, in a particularly conspicuous manner compared to other new features.

Our initial goal with this work was to document and analyse the design strategies employed by tech companies in order to bring about the announced \textit{AI revolution} \cite{makridakisForthcomingArtificialIntelligence2017}. While new digital services were designed as components of this "revolution", we chose to focus on the strategies that aim at favoring user engagement with AI-based features on digital platforms that were initially not conceived as AI-first products.

Having started to collect extensive anecdotal evidence on the reluctance of users feeling pressured to use AI-based features in their daily tools, we also want to understand to what extent these design strategies fall within existing deceptive design patterns categories.
In doing so, we could build on Tocze et al.’s work on unsustainable patterns \cite{toczeDarkSideCloud2022} by describing the ways in which design patterns can favor or force the adoption of digital uses that are environmentally unsustainable, for the benefit of companies. 
Our hope for this work is that it will spark debate on unsustainable use adoption and on the opportunities and challenges involved in the regulation of the forced adoption of unsustainable digital products and services.

%% file: tex/2_background.tex
\section{Background}
While Artificial Intelligence (AI) has always been a subject of debate, recent developments have stoked controversies surrounding biases, misinformation and privacy issues \cite{baldassarreSocialImpactGenerative2023}. Our work builds upon research on the environmental effects of AI and explores the challenges involved in exposing the role of design in intensifying demand and related indirect effects.

\subsection{Environmental effects of generative AI}
Research have been denouncing the environmental costs of large-scale AI development for years \cite{sashaluccioniEstimatingCarbonFootprint2022, strubellEnergyPolicyConsiderations2020, morandHowGreenCan2024}. The impact extends beyond climate, affecting 6 out of 9 planetary boundaries\cite{falkAttributionProblemSeemingly2024}; additionally the environmental benefits of \textit{AI for green} solutions are now called into question \cite{ligozatUnravelingHiddenEnvironmental2022, morandHowGreenCan2024}. Yet, generative AI services remain based on energy-intensive architectures \cite{luccioniPowerHungryProcessing2024}. Also, AI systems and products are produced in a context that favors constant scaling up to match increasing computational possibilities. Even in AI research, a \textit{"bigger-is-better"} narrative perpetuates the assumption that the bigger the AI system, the better its performance and value~\cite{varoquauxHypeSustainabilityPrice2025}. 

A recent study led between September 2023 and August 2024 on the electricity consumption of data centers found they account for over than 4\% of the US's electricity consumption, with 56\% of these originating from fossil fuels \cite{guidiEnvironmentalBurdenUnited2024}. The authors also project that the \textit{AI race} will play a role in the doubling of the worldwide energy consumption of data centers, which would outgrow the power used in Canada (the sixth highest energy-consuming country in the world) by 2026. As they put it, \textit{"the AI race significantly increases computational demands and, consequently, energy consumption of data centers"} \cite{guidiEnvironmentalBurdenUnited2024}.
While little data exists on the use phase of large models, Varoquaux et al. \cite{varoquauxHypeSustainabilityPrice2025} have estimated that the energy use due to inference (i.e. the application phase of AI) would outgrow the training phase amount of energy use only in a few weeks in the case of ChatGPT. Greenhouse gas emissions due to inference are slated to worsen the already alarming impact of the biggest tech companies: in 2023, Google’s emissions increased by 48\% compared to 2019 and Microsoft's increased by 29\% compared to 2020 \cite{guidiEnvironmentalBurdenUnited2024}.

Luccioni et al. 
\cite{luccioniEfficiencyGainsRebound2025} point to the risks of rebound effects if AI development keeps on following an exponential growth trend. 
They also note that while the power usage effectiveness (PUE) of data centers training AI is improving, their global energy use is on the rise. 
This rebound effect is tightly connected with increased AI use due to an expanding market, the impact of which cannot yet be fully assessed. This relates to the cornucopian paradigm \cite{preistUnderstandingMitigatingEffects2016}, which describes how demand is stimulated by the design of digital services, which in turn drive the expansion of the digital infrastructure, e.g. manufacturing more GPU chips and building new hyperscale data-centers. 
In this paper, we seek to document and analyse the strategies deployed by companies to stimulate AI adoption and ultimately lead to rebound effects.

\subsection{Deceptive Patterns}
As gen-AI is a controversial technology, the question of user agency regarding its adoption is important. 
Since 2010, dark, deceptive and manipulative design patterns have been developed, and have ended up  "limit[ing] user autonomy and decision making" \cite{grayOntologyDarkPatterns2024}.

Ibrahim et al. \cite{ibrahimCharacterizingModelingHarms2024} have explored both the individual and social implications of the design of AI interface features. They discuss the \textbf{"cascading impacts"} of AI use when deceptive design choices lead to addictive behaviors or encourage the spread of misinformation. Our study builds on this work by focusing specifically on the design strategies used to push for AI adoption, to connect features' design with environmental concerns.

While we argue that certain design choices are harmful, we do not argue that they are deliberately harmful. In this sense, we align with Di Geronimo et al.’s \cite{digeronimoUIDarkPatterns2020} approach of deceptive patterns which analyses when an \textit{"interface seem[s] to benefit the app rather than the use"}. We focus on interface choices that have benefited generative AI use, understood here as an environmentally problematic technology, and identify design strategies that have not been conducive to users’ autonomy, freedom of choice and understanding.
We adopt a \textit{collective welfare lens} \cite{mathurWhatMakesDark2021a} to study deceptive design patterns, meaning that we focus less on individual consumer rights and more on socio-technical transformations liable to deteriorate living conditions on earth. 

A previous LIMIT publication \cite{toczeDarkSideCloud2022} coined the term \textit{"unsustainable design patterns"} which describes how the arrangement of different components (e.g. \textit{"artifact", "intent", "participant", "context"}) results in unsustainable shortcomings. 
Our study focuses on the \textit{"artifact"} as a means to highlight potential unsustainable design patterns. Our work expands in particular on \textit{"planned rebound effect"} pattern, whereby platform providers would be \textit{"pretending to ignore what could be a consequence of launching a new service"}. Our analysis uncovers how this pretending strategy is embedded in interface choices in the case of AI-based features in 2024. 
We link this analysis to previous work using Gray et al's \cite{grayOntologyDarkPatterns2024} ontology of patterns, which aims for a \textit{"shared language"} between legislative and academic communities. We reflect and speculate on how design strategies leading to imposing AI use could be denounced and eventually regulated.

%% file: tex/3_method.tex
\section{Method}
In early 2024, we started noticing that many AI icons were appearing in the software and applications we use. From September 2024 onwards, we started, as a team of 3 HCI researchers, to systematically collect screenshots and descriptions of each new AI-related UI or UX change we encountered. We also collected and documented cases that we came across on social media.

To diversify our corpus, we first asked for input from colleagues, family and friends through word of mouth. We also launched open calls on Mastodon via the Limites Numériques account, which has over 1,000 followers\footnote{https://mastodon.design/@limitesnumeriques/113431042717320191}. Over the course of a 4 month period, we collected \textbf{90} screenshots and descriptions of recent AI related changes in interfaces in \textbf{53} different software and app interfaces.

We only collected examples explicitly labeled or presented as “AI”. We also only focused on software and application that previously operated without generative AI-based features. We should note that our corpus is not representative of the state of global software usage and reflects our bias towards the applications we use on a daily basis, which were mainly developed in the US and to a lesser extent in Europe. However, in this study we focus on identifying and analysing the strategies used to favor AI and not on such representativeness issues, which is why we do not provide counts or percentages when describing our findings.

Our corpus includes many of the biggest software companies, including Google, Apple, Meta, Microsoft and Adobe. We found examples in both software geared towards professionals such as Slack or Adobe Photoshop, but also numerous examples of leisure, entertainment or social media applications such as Spotify, Microsoft Paint, Instagram or Snapchat. To put our corpus in perspective, we complemented our analysis with written sources such as press releases and ads targeting a general audience but also more technical communications (e.g. companies’ blog posts) intended for designers and developers.

We constructed our method of analysis in two steps. The three authors first collectively analysed the corpus using a clustering technique. We grouped examples that were using the same strategies together and coded these strategies with tags such as "takes up space", "magic", "unremovable", etc.). As strategies started to consolidate, we went back to the corpus and tagged all the corpus items with these strategies to make sure that we were able to reliably identify which strategies were being used. From this analysis, we observed recurring strategies in terms of wording, information architecture and graphic design choices, as well as interaction patterns.

Based on this initial analysis, we wanted to verify whether and to what extent these recurring patterns were consistent with the ontology of existing deceptive patterns. We chose to use Gray's comprehensive ontology \cite{grayOntologyDarkPatterns2024} to map the specificities of the patterns we encountered.

%% file: tex/4_results.tex
\section{Results}

\begin{figure} [ht]
    \vspace{-.2cm}
    \centering
    \includegraphics[width=1\linewidth]{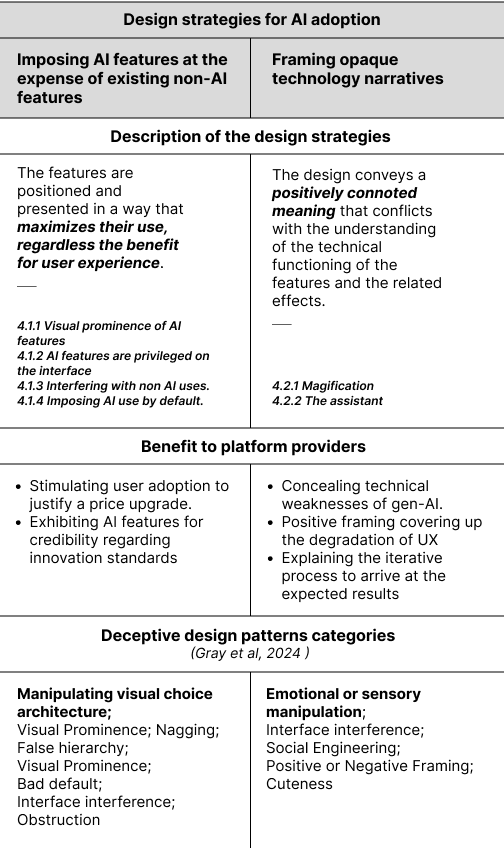}
    \vspace{-.3cm}
    \caption{Summary table of our two strategies and the related deceptive patterns}
    \label{fig:Tableau}
    \vspace{-.3cm}
\end{figure}

Based on our analysis, we evidenced two types of strategies that rely on a combination of deceptive patterns. We first discuss the visual and interaction design strategies that affect the expected affordances (4.1), information architecture and user flows in order to favor AI adoption, generally at the expense of existing digital practices on the application or platform. We then turn to a second more insidious strategy that relies on the manipulation of narratives around technology use through interface design choices (4.2).
Relying on the analysis of interface elements, we used Gray et al's ontology of deceptive patterns ~\cite{grayOntologyDarkPatterns2024} to situate identified  strategies in relation to known deceptive patterns. Our analysis highlights the use of two specific meso-level deceptive patterns~ (Figure \ref{fig:Tableau}): \textit{"Manipulating visual architecture"} and \textit{"Emotional and Sensory Manipulation"}. 
In our case, however, it is the intersection and combination of manipulative tactics across many widely used platforms that really gives us insights on how AI use is imposed to users. 
These strategies also all align with the \textit{``planned rebound effect"} pattern from Toczé et al.~\cite{toczeDarkSideCloud2022}, as the wide scale adoption of AI technology will increase the ICT's sector footprint, as well as with \textit{``Environmental blindness"} and \textit{``concealed impacts"}, as the strategies employed are making AI's environmental impact less visible.

\subsection {Imposing AI features at the expense of existing non-AI features}

As Janlert and Stolterman have shown \cite{janlertThingsThatKeep2017}, visual interfaces are finite space, i.e. a bottleneck, which means designers have to make choices about what features to display and how much space can be devoted to each of them. In our analysis we found that, in comparison with non-AI features, AI-based features tend to occupy a place of honor and to call for users’ attention in many different ways. AI-based features are positioned and presented in a way that maximizes their use, regardless of whether this benefits the user experience.

\begin{figure} [ht]
    \vspace{-.2cm}
    \centering
    \includegraphics[width=1\linewidth]{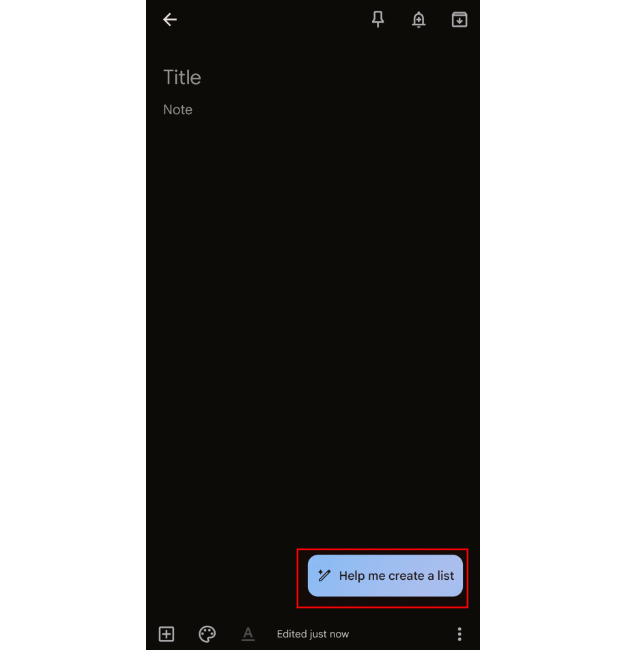}
    \vspace{-.2cm}
    \caption{Floating button for an list generation feature on Google Keep}
    \label{fig:googlekeep}
    \vspace{-.1cm}
\end{figure}

\subsubsection{\textbf{Visual prominence of AI features}}
\begin{figure} [ht]
    \vspace{-.2cm}
    \centering
    \includegraphics[width=1\linewidth]{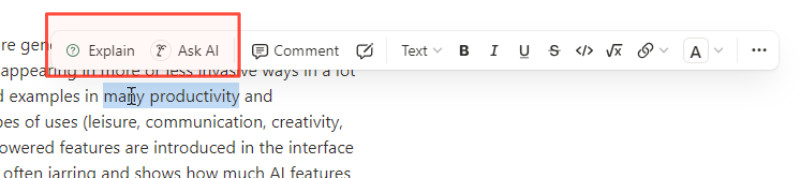}
    \vspace{-.3cm}
    \caption{AI-related features are the first on Notion's toolbar takes up a the third of the space}
    \label{fig:notion}
    \vspace{-.1cm}
\end{figure}

AI features tend to take up a lot of space and are often given the most valuable space in interfaces, toolbars and menus. Linkedin’s interface, for example, features a messaging popover and a banner advertising an AI-based feature that occupy more than half of the space. In Notion, AI features take up nearly a third of the toolbar (Figure \ref{fig:notion}). In both Notion and DeepL, whenever users select words, a floating toolbar appears in which the buttons closest to the cursor are AI features. This indicates that they are treated as the most important features.

Interfaces often feature one salient call to action and these are also being used to encourage the use of AI-based features. In Google Keep, a floating action button, which is the component that helps people take primary actions, displays an AI-powered feature (Figure \ref{fig:googlekeep}). Similarly, Google’s Gemini is embedded by default on all Galaxy S25 devices and can be activated by a long press on the power button.

\begin{figure} [ht]
    \vspace{-.2cm}
    \centering
    \includegraphics[width=.9\linewidth]{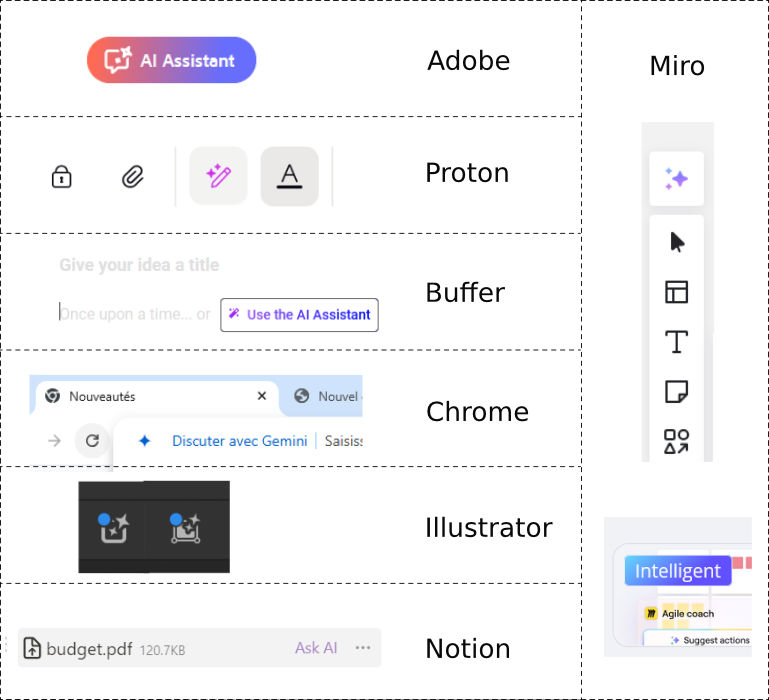}
    \vspace{-.3cm}
    \caption{Icons of AI-based features, enhanced through graphic design.}
    \label{fig:icons}
    \vspace{-.3cm}
\end{figure}

\begin{figure} [ht]
    \vspace{-.2cm}
    \centering
    \includegraphics[width=.9\linewidth]{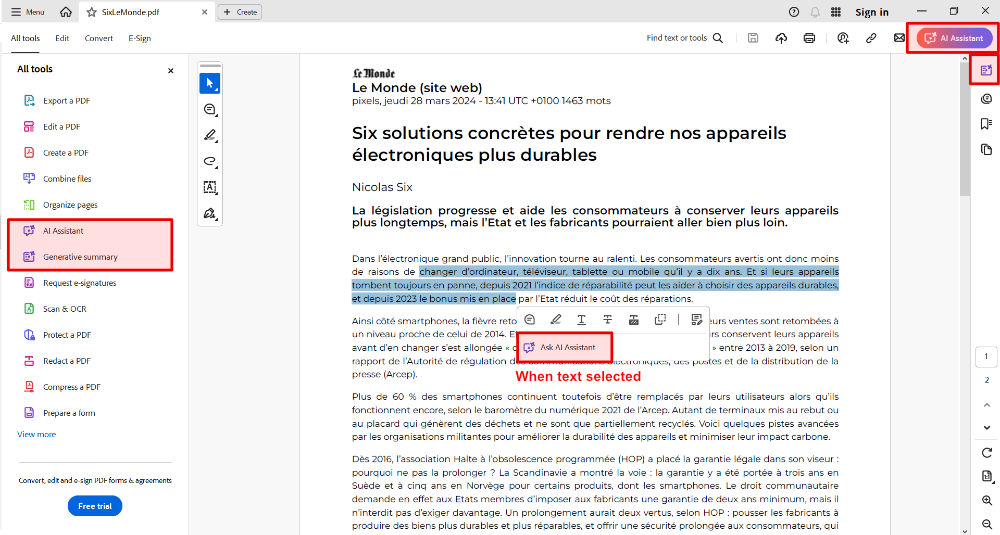}
    \vspace{-.3cm}
    \caption{Adobe acrobat reader interface, where AI features are redundant.}
    \label{fig:Adobe}
    \vspace{-.3cm}
\end{figure}

\begin{figure} [ht]
    \vspace{-.2cm}
    \centering
    \includegraphics[width=1\linewidth]{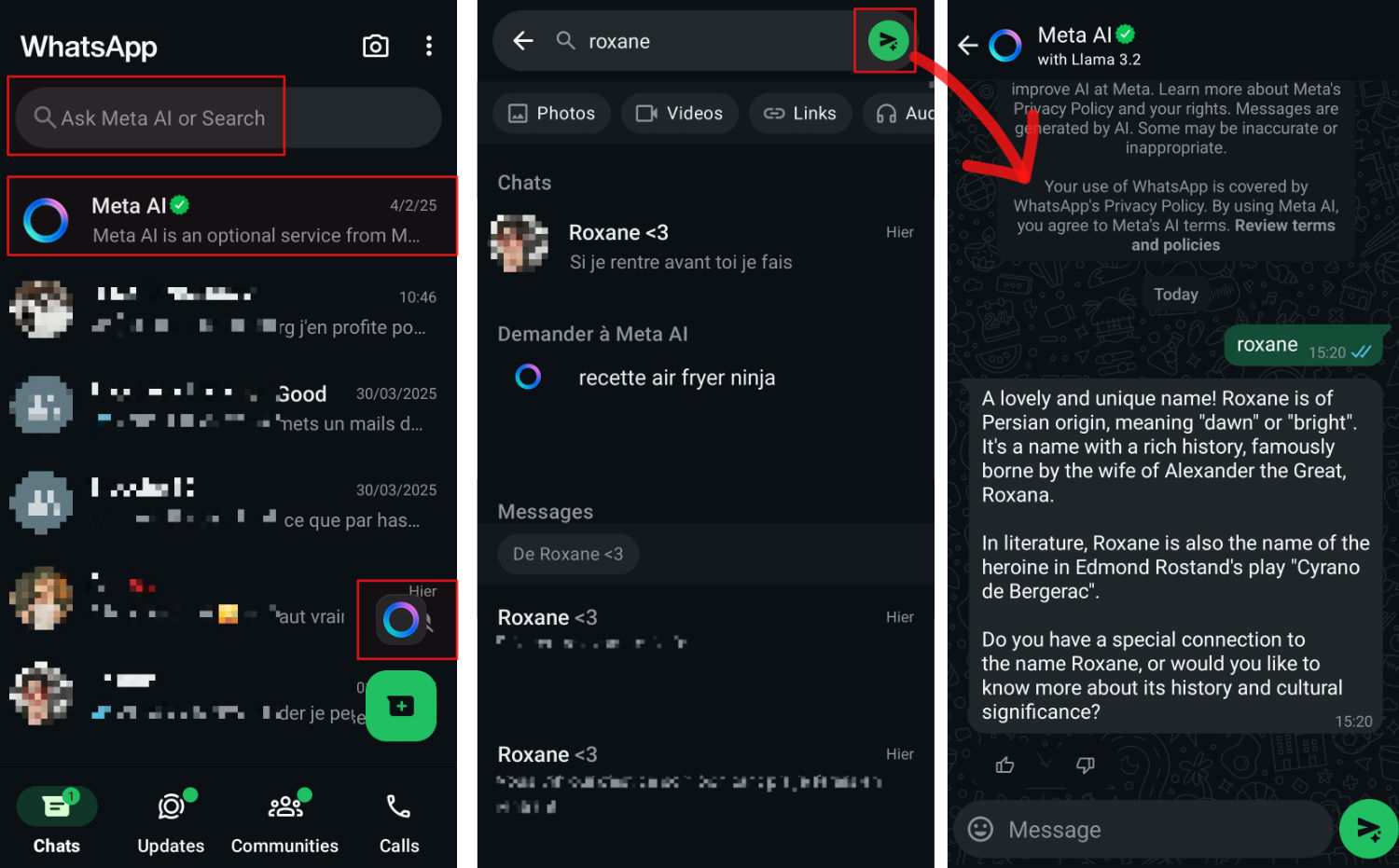}
    \vspace{-.3cm}
    \caption{WhatsApp interface: the search bar has deviated from its initial purpose and can now be used for prompt inputs.}
    \label{fig:Whatsapp}
    \vspace{-.3cm}
\end{figure}

 AI features stand out visually as well as spatially. In numerous cases (Figure \ref{fig:icons}), AI-based features are highlighted using a distinctive color, a colorful gradient , or even using an animated icon (Notion) in contrast with all the other features, which are usually displayed using static black and white icons. We also found that AI features are given multiple access paths and that users are constantly reminded about these features. In Acrobat Reader, for example, AI features are presented and suggested to users on multiple occasions (Figure~\ref{fig:Adobe}).  The same “AI assistant” feature is presented in both left and right toolbars, as a button in the menu bar, as a tooltip in the interface, as a GUI prompt and also when users select text. Similarly, Whatsapp has updated its applications to integrate even more AI-based features (Figure  \ref{fig:Whatsapp}). Besides the addition of a "Meta AI" chat feature, the search bar originally dedicated to retrieving contacts and conversations now also redirects to Meta’s AI assistant. 


\begin{figure} [ht]
    \vspace{-.2cm}
    \centering
    \includegraphics[width=1\linewidth]{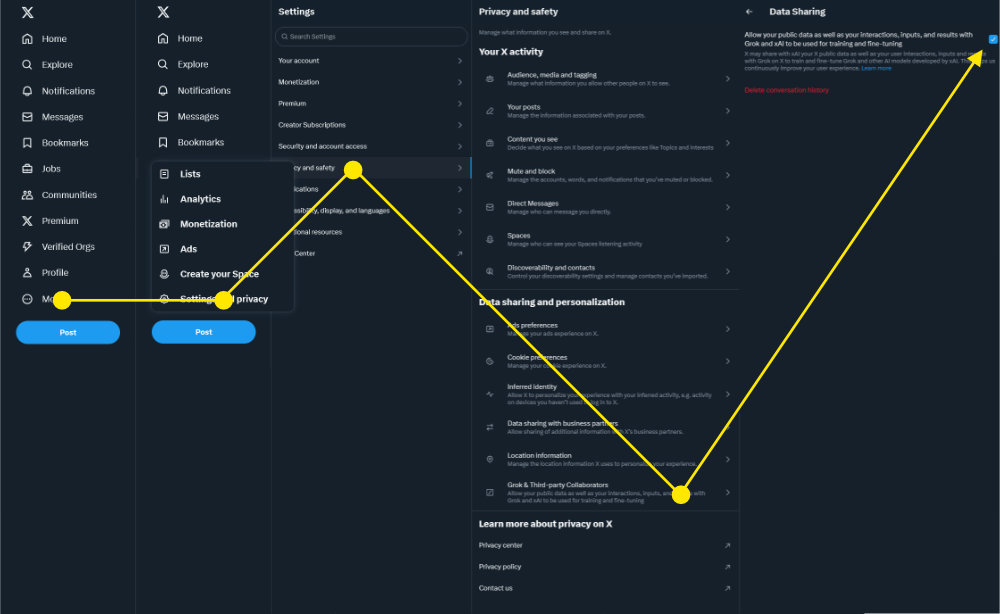}
    \vspace{-.3cm}
    \caption{The path for deactivating personal data training on X involves 5 steps}
    \label{fig:optout-x}
    \vspace{-.3cm}
\end{figure}

This visual prominence of these examples of AI features contrasts with the discretion that is exercised when it comes to how user data is being used to train AI models on many applications. If the option to disable such personal data use exists, its existence is not notified to users and the dedicated parameter is very often hidden among others (Figure  \ref{fig:optout-x}).

These design strategies are a good match for Gray et al.'s \cite{grayOntologyDarkPatterns2024} \textit{"Visual Prominence"} pattern, which places \textit{"an element relevant to user goals in visual competition with a more distracting and prominent element"}. 
This omnipresence is even reinforced with strategies consistent with the pattern \textit{"nagging"} \cite{grayOntologyDarkPatterns2024}, where AI features are presented repeatedly and interrupt users in their actions. 
For example, when opening Adobe Photoshop, users are first welcomed with a popup encouraging them to \textit{'explore the power of generative AI'}. After closing this initial popup, a tooltip opens in the interface to suggest the same feature. Similarly, when opening Microsoft Skype, users were often greeted by its AI-chatbot Copilot, which invited users to use it as soon as the interface opened, regardless of whether they wanted to use that feature or not. These strategies may induce users who were originally uninterested in using AI to click on the features out of curiosity or by mistake. This pervasive advertising of AI also leverages the perception that AI use is an unavoidable technological turn. 

\subsubsection{\textbf{AI features are privileged on the interface}}
\begin{figure} [ht]
    \vspace{-.2cm}
    \centering
    \includegraphics[width=1\linewidth]{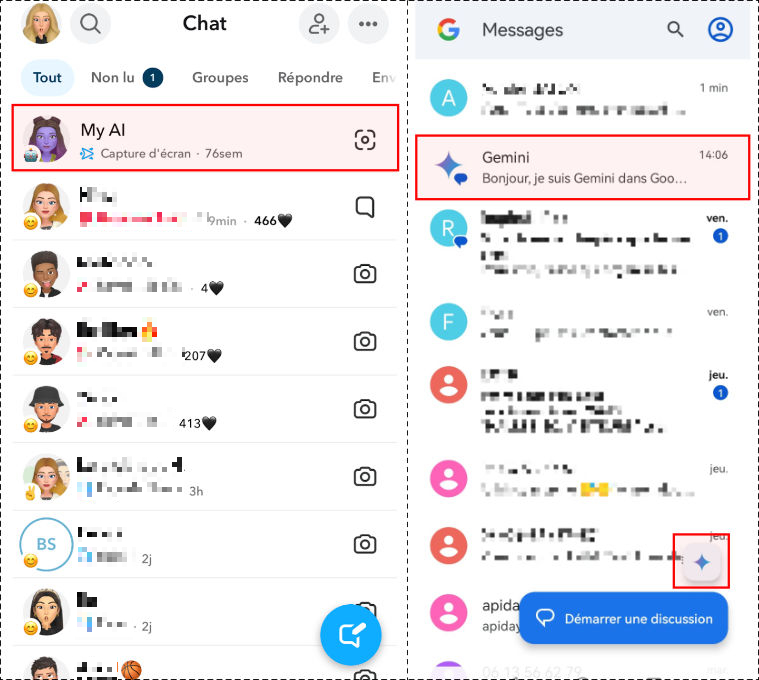}
    \vspace{-.3cm}
    \caption{AI assistants on Snapchat and Messages, displayed as contacts, regardless of whether users interact with the assistant.}
    \label{fig:MyAI}
    \vspace{-.3cm}
\end{figure}

We noticed that in some cases, the strategies go beyond making AI features visually prominent. We found examples of AI features that are presented in the same fashion as other features or other interface elements through their visual organization and form. AI assistant features are implemented on messaging apps in the form of a conversation, meaning that assistants use the same UI elements as regular discussions with contacts (Figure \ref{fig:MyAI}). On Snapchat, on the other hand, the AI assistant, unlike all the other conversations, always remains on top and cannot be removed, regardless of whether it is actually used or not. While other conversations are gradually dropped down if they are never updated, AI assistants are not subject to the same treatment and limit the screen space dedicated to active conversations. The hierarchy rules that apply to similar UI elements do not apply to AI features, in a way that advantages the visibility and the interaction with the feature. Similarly, as noted earlier, AI features are sometimes highlighted through the presence of multiple buttons on the screen, where the other features generally only have one (Figure \ref{fig:Adobe}).

These examples match Gray et al.'s \cite{grayOntologyDarkPatterns2024} \textit{"False Hierarchy"} pattern where the strategy is to \textit{"give one or more options visual or interactive prominence over others, particularly where items should be in parallel rather than hierarchical"}. This \textit{"manipulation of visual choice architecture"} influences the choice of features to interact with: AI-based features are discretely favored among others using UI/UX design.

\subsubsection{\textbf{Interfering with non AI uses}}
We found that the ways AI features are pushed on interface users also interfere with the use of non-AI features and interrupt users’ usual workflows. Many tools advertise their AI features using banners in the interface, including Slack and Google Docs. The tooltips and banners promoting AI in interfaces generally require at least a click from users to dismiss them. Moreover, the preeminence given to AI features inevitably leads to people triggering AI features, sometimes accidentally. For example, in Notion, it is very easy to launch the AI assistant inadvertently, since  pressing the extremely frequently used space bar will open it. In comparison, the keyboard shortcut for any other command requires users to press ‘/’, which is less conducive to this type of error.

A similar hijacking of user experience on the platform happens when actions routinely performed before the AI feature update take more time and become more fastidious. On Qwant, the results of a web search engine have been downgraded, now appearing below a huge box of AI-generated responses, requiring users to scroll longer. 

\begin{figure} [ht]
    \vspace{-.2cm}
    \centering
    \includegraphics[width=1\linewidth]{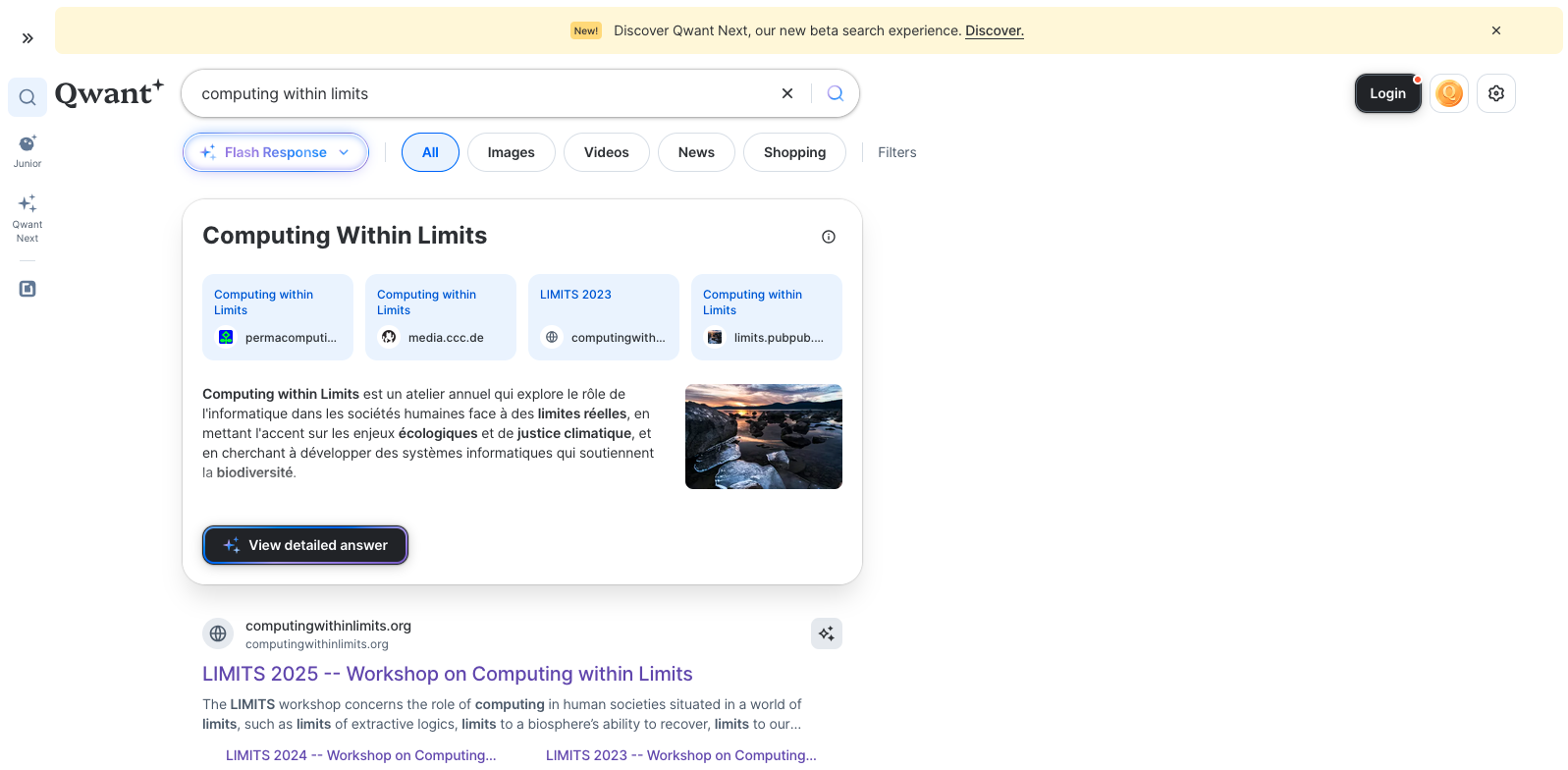}
    \vspace{-.3cm}
    \caption{Qwant search results.}
    \label{fig:Qwant}
    \vspace{-.3cm}
\end{figure}

These interferences with the use of non-AI features present significant similarities with Gray et al.’s deceptive pattern \textit{"Adding Steps"}. Even if it has so far mostly been used to describe privacy related issues, \textit{"Adding Steps"} to perform a given online task is a deceptive pattern that is arguably very relevant in contexts in which a company has an interest in influencing users’ adoption of features. In the context of AI, this type of obstruction strategy disrupts users’ habits on the platforms. Users who do not want to use AI will have to tolerate this friction on their existing digital habits, or to look for a way to disable -if possible- the feature that is bothering them.

\subsubsection{\textbf{Imposing AI use by default}}

As we have seen, AI features are extremely easy to trigger, but they are also sometimes enabled by default or deployed before users are allowed to choose to deactivate them.

\begin{figure} [ht]
    \vspace{-.2cm}
    \centering
    \includegraphics[width=1\linewidth]{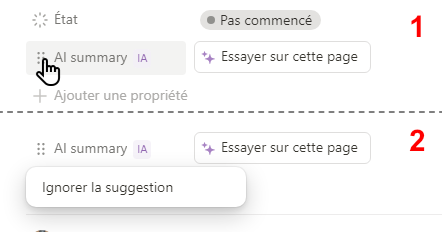}
    \vspace{-.3cm}
    \caption{Notion database implementing a AI summary field by default}
    \label{fig:ainotiontable}
    \vspace{-.3cm}
\end{figure}

\begin{figure} [ht]
    \vspace{-.2cm}
    \centering
    \includegraphics[width=1\linewidth]{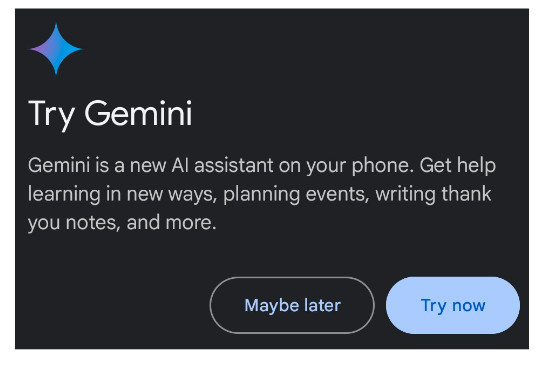}
    \vspace{-.3cm}
    \caption{On Google Play, turning off an AI-based feature is not a permanent decision.}
    \label{fig:Notnow}
    \vspace{-.3cm}
\end{figure}

For example, the sport tracking application Strava introduced AI comments on activities without offering users the option to turn them off. YouTube also started to automatically dub videos. Users can turn it off in each individual video but to our knowledge there is no option to disable it entirely. In Notion, an \textit{‘AI Summary'} property is featured by default when creating a new database, whereas many other properties are not (Figure \ref{fig:ainotiontable}). The AI property requires  explicit action from users looking to delete it. Unlike all the other properties, which can be deleted using the  \textit{‘delete property'} menu option, the AI summary property can only be dismissed using the \textit{‘ignore property'} menu option. Even when applications provide ways to disable AI-based features, they involve friction. This may first be a result of the wording, as refusing to use AI based feature can rarely be a firm no. Instead, applications often offer to turn off the feature temporarily, using words such as \textit{‘ignore for the moment'} or \textit{‘maybe later'} (Fig: \ref{fig:Notnow}). What this tells users is that they will certainly come to use it in the long run.

The default activation of AI features, making it much harder for users to deactivate or altogether remove them from the interface, matches Gray’s pattern category \textit{"Bad Defaults"} \cite{grayOntologyDarkPatterns2024}, where default settings are not in the users’ best interest; in this case, these default settings also go against sustainable digital practices, which is arguably against the long-term best interests of users (and non-users).

\subsection{Framing opaque technology narratives}
Having documented a first strategy to push AI-based features that consists in making them stand out on the interface and/or interfere with non-AI user experience, in the following we expand on two technology narratives that are distinct in form but have similar effects, related respectively to the covering up of technical weaknesses and the controversies around artificial intelligence. 

\subsubsection{\textbf{Magification of AI technicality}}
The magic metaphor has long been associated with the use of computers and more globally to innovation \cite{kellyWorksMagicMetaphor2018}. In 2007, the first Iphone's Multi-Touch system was already \textit{"working like magic"} in the words of Steve Job \footnote{See the video of the keynote here: https://www.youtube.com/watch?v=roTRqbhcpyQ}. Already in 1989, Photoshop's magic wand icon allowed communicating users the automatic nature of the selection process of an element \footnote{Other historical examples of magic metaphors in interaction design can be found in Brulé's blog article : https://sociodesign.hypotheses.org/942}. However, we observe the magic metaphor is now diffusing at an unprecedented level on interfaces. The omnipresence of the magic metaphor in the ways AI features are presented is unusual enough to wonder how do narratives based on magic benefit tech companies. We rely on Nagy and Neffs' \cite{nagyConjuringAlgorithmsUnderstanding2024} work on the deployment of magic narratives by the tech industry, and investigate the particular role of design in the "\textit{conjuration of algorithms}", i.e. \textit{"a strategically deployed narrative device that uses the principles of magic to manipulate the public perception of technologies"}.
Lupetti et al. \cite{lupettiUnmakingAIMagic2024} for example, have shown how the magic metaphor can be used as a way to control the social meanings embedded in technology use: 
\begin{quote}
\textit{"Designing and communicating AI things as supernatural–enchanted–products, in fact, shapes the social perceptions of these systems, taking them out of the realm of mere technical tools to be regarded as socially capable agents [...] and/or socially valuable applications"}
\end{quote} 

In user interfaces as well, AI is represented as magic to favour its adoption. The most common graphic symbol for representing an AI-based feature is the spark icon. Generally associated with something special, exciting, new, but also with innovation and wonder, the icon contributes to a representation of an intrinsically \textit{good} functionality. Unlike other icons that literally or metaphorically illustrate features, the icon used for AI sets it apart from other interface features. Likewise, mauve-centric gradients, commonly associated with magic and the impalpable, are predominantly used for AI.

\begin{figure} [ht]
    \vspace{-.2cm}
    \centering
    \includegraphics[width=1\linewidth]{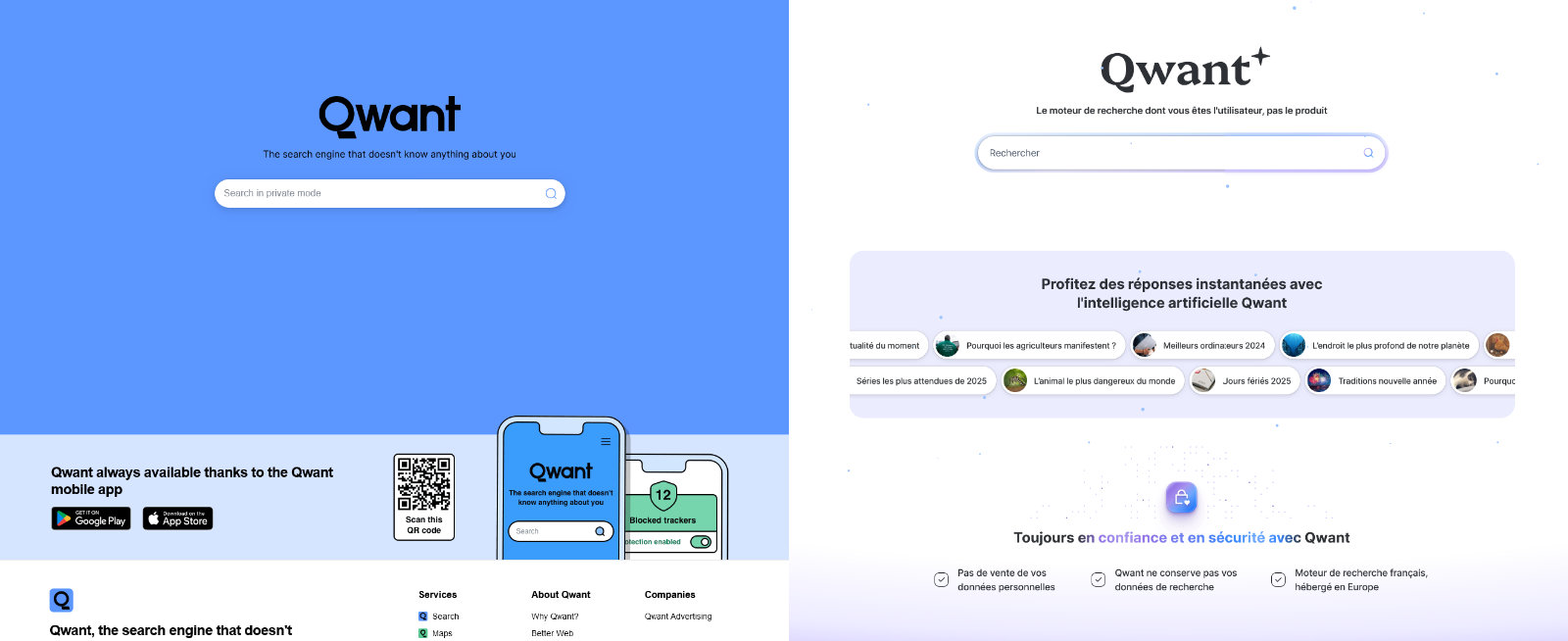}
    \vspace{-.3cm}
    \caption{Qwant's home page in April 2024 (on the left) and November 2024 (on the right). On the right screen, a baseline introduces the new AI features: "Enjoy instantaneous answers with Qwant's artificial intelligence" (Translated from French)}
    \label{fig:Qwantmagic}
    \vspace{-.3cm}
\end{figure}

Qwant offers by default what it calls an \textit{‘flash’} response to searches thanks to a LLM feature. Ironically, this \textit{‘flash’} response actually appears long after the web results. The words and metaphors used here make the machine’s calculations invisible, concealing the fact that it is much slower than a traditional search engine and much more resource-intensive. We don’t know how something magical should behave: is it normal for it to take time? Have I done something wrong? Is its consumption of resources justified? The invisibility of the machine enabled by the metaphor of magic puts people in a different relationship to the tool, in a way that diverts attention from its weaknesses – here, the increased response delay. It is no longer possible to pay attention to the machine. While metaphors are useful to the wide-scale appropriation of complex and new technologies, previous work has shown that digital metaphors tend to have an impact on the technological culture, and in turn normalize digital behaviors associated with a limitless, non-impacted, \textit{"immaterial"} digital infrastructure \cite{borningInvisibleMaterialityInformation2020, borningWhatPushesBack2018}.

By revising their graphic identity using the visual semantics of magic (Figure \ref{fig:Qwantmagic}), companies position AI as an all-purpose tool, without ever saying what it does not do. They benefit from setting such vague expectations among users, as it is more difficult to criticize the results of an action that had no precise goal in the first place, in the same way that surprise is the expected outcome of a magic trick. The promised versatility of AI tools, embodied in magical evocations, also results in the invisibilization of the materiality of the many operations involved in AI, spreading the idea that AI necessarily makes things easier, better and faster.

Interestingly, both magic and deceptive patterns can be closely connected to trickery. The language of magic conjures a sense of enchantment, i.e., \textit{“the experience of being caught up and carried away”} \cite{bennettEnchantmentModernLife2016}, where knowing what happens behind the curtain is not only unnecessary, but also detrimental to the enjoyment of the full \textit{magical} experience. Such metaphors allow to sidestep questions of efficiency and environmental or political effects, framing an opaque process of data generation into a magical event.

\subsubsection{\textbf{The dedicated assistant}} 
Another recurring form taken by generative AI in our corpus is the assistant.

The assistants offer to help people \textit{‘learn new ways, plan events, write thank you notes, and more’} (Gemini on Google), \textit{‘chat, create and find anything’} (Notion), or \textit{‘suggest unique ideas’} (Skype). 
AI assistants are embodied using metaphors that evoke human characteristics, for example given names like ‘Aria’ (Opera) or ‘Leo’ (Brave), and are also assessed through the same UI elements as human collaborators or contacts (Figure \ref{fig:MyAI}). 
The tradition of attributing human characteristics to machines and robots has already been discussed in 1986 by Caporael \cite{caporaelAnthropomorphismMechanomorphismTwo1986}, who unpacked the cognitive biaises behind the assumption that a machine with human attributes will perform better. Deploying particular metaphors of human roles or human characteristics leverages \textit{"previously known mental schemas for social relationships"}\cite{newendorpApplesKnowledgeNavigator2024}, meaning that it shapes users' expectations of technology and thus the way they will approach it. 
In our case, presenting AI as having human characteristics or skills contributes to the discourse presenting AI as desirable and high-performing. This emphasis on versatile human skills also helps to justify the introduction of AI features into our digital services: AI can do everything, so it can be everywhere.

Generative AIs’ names often evoke a role in the professional environment, such as  \textit{‘Copilot'} (Microsoft), \textit{‘Agentforce'} (Slack), \textit{‘Cocreator'} (Paint), \textit{‘Companion'} (Microsoft, Zoom), \textit{‘Analyst'} (Google Chat) and, of course, \textit{‘Assistant'} (Protonmail, Brave, Webex). These names evoke collaboration among peers, sometimes represented by a personified emoji (Notion). They also echo the fantasy of alleviating demanding tasks through technology, a narrative already pushed in Apple's 1987 keynote \cite{appleinc.EDUCOM1987John1987}: thanks to the AI assistant, uninteresting or difficult tasks can be delegated, allowing users to focus on more important duties. The sentences that present the integration of AI assistants generally give an impression of availability and dedication (ex: \textit{‘Whenever you need me'} (Notion), ‘Need help?' (Brave).

AI assistants are presented as polyvalent subordinates, available for any task that can be formulated through a textual prompt. AIs are positioned under us in the hierarchy in a context of fear of labor degradation for skilled workers. Its polite and \textit{"discreet"}\cite{bennettEnchantmentModernLife2016} demeanor also reassures users about the safety of their personal data. Such  \textit{"undue trust"} in a robot is a deceptive pattern referred by Gray et al. as \textit{"cuteness"}. It conveys the appealing figure of an agent that combines the properties of a tool and of a teammate, leaning towards the \textit{"agent-as-a-teammate paradigm”} \cite{newendorpApplesKnowledgeNavigator2024}, where \textit{“we want tireless agents that help us while treating us like teammates, even if we don’t treat them like teammates”}. 

This second strategy (4.2) is distinct as it is part of global narratives that have been disseminated beyond platforms (e.g. in commercials, on promotional websites) for decades \cite{newendorpApplesKnowledgeNavigator2024}. Among other effects, these metaphors contribute to framing a competitive climate where AI use equals professional performance \cite{pasquinelliAIWorkAutomation2024}. In an environment where AI is presented as unbiased, more efficient and unavoidable, managers and employees might be more prone to adopt useless AI-based features out of a fear of appearing outdated or incompetent.

This impact at the social scale links to \textit{"social engineering"} deceptive patterns \cite{grayOntologyDarkPatterns2024}, where design choices influence perceived norms around technology use. One important shared trait of the magic and assistant metaphors is that they disseminate unrealistic expectations of technology to favor its blanket adoption in multiple sectors.

Both of the metaphors align well with Gray et al.'s design pattern \textit{"emotional or sensory manipulation"} \cite{grayOntologyDarkPatterns2024}, where design choices are used to \textit{"evoke an emotion or manipulate the senses in order to persuade the user into a particular action"}. Metaphors contribute in creating the nature of the emotional relation between users and interactive systems : according to Murray \cite{murrayInventingMediumPrinciples2012}, a magical relation places users in trusting that the system knows his intentions and next actions, instead of  making users expect to control all aspects of the system for example. 
Natale has shown how analogies used in the scientific literature contribute to shaping the \textit{"AI myth"} \cite{nataleImaginingThinkingMachine2020}; likewise, design choices for AI features contribute to the construction of technological myths. 
In the case of AI, these narratives are strongly connected to ongoing debates on the weaknesses and concerns around artificial intelligence. For example, both metaphors are intended to mitigate  \textit{"algorithm aversion"} \cite{casteloTaskDependentAlgorithmAversion2019}, i.e. the fact that users value more human abilities than AI’s. Indeed, magic blurs and positively frames the outcomes of AI-mediated operations, while the assistant metaphor suggests a quality equivalent to human-produced content. 

Conversely and more surprisingly, these metaphors also put users in a position to expect inaccurate results from AI features. The magic metaphor explains approximations in AI responses and encourages users to make other tries by adding some tweaks until they find the right \textit{trick}. Comparably, the assistant has to be briefed \textit{correctly} to function, meaning the responsibility of AI interaction’s outcome falls on users, who need to adjust their prompts to steer the assistant in the right direction. In both ways, the positive framing of these narratives is combined with an indication that the AI feature might not immediately work as we would like it to.

%% file: tex/5_discussion.tex
\section{Discussion}
While the design strategies used to favour AI adoption do fit within the existing deceptive pattern categories described by Gray et al. \cite{grayOntologyDarkPatterns2024},
Brignul's initial overall definition of deceptive patterns (\textit{"tricks used in websites and apps that make you do things that you didn't mean to, like buying or signing up for something"}\footnote{https://www.deceptive.design/}) falls short in our case. Here the patterns are not deceptive \textit{per se} : users know that they are using AI features and no immediate damage occurs, like an unintended purchase. The issue here is that use is provoked by leveraging power dynamics that are beyond any user’s agency. In the case of AI’s deceptive patterns, the tricks are also performed on another temporality and on another scale, reminiscent of Toczé et al.’s design pattern \textit{"planned rebound effects"} \cite{toczeDarkSideCloud2022}. 
Chordia et al. \cite{chordiaDeceptiveDesignPatterns2023} have coined the term \textit{"deceptive infrastructure"} as a way to present the interaction of socio-economical factors and power dynamics with interface features with a negative impact on the user. The deceptive patterns identified in this study similarly interact with broader economic and environmental contexts, which results in long-term systemic impacts rather than in instantaneous individual damage. Because the end goal of these patterns is not deception but enforced adoption, design strategies aimed at imposing the domination of unsustainable digital features on the market need to be better characterized, so that in turn we know on which grounds they could be regulated.

We have relied on the concept of deceptive patterns to analyze our findings because it has already been used to bridge design and regulation actors \cite{grayOntologyDarkPatterns2024}. 
Regulating AI deployment on the basis of its environmental impact poses a challenge, as do all environmental regulations, which is why different opportunities for regulating the imposed adoption of AI are suggested here. Deceptive patterns could be used as a way to support regulations on a variety of dimensions that are key for reducing the environmental footprint of the ICT sector more generally: connecting sustainability interests with fair market competition; favoring individual sustainable practices; and promoting environmental transparency.

\subsection{Growth, market capture and lock-ins}
\begin{quote}
    \textit{"Attempts to present AI as desirable, inevitable, and as a more stable concept than it actually is follows well-worn historical patterns: one of the most important ways for a technology to gain market share and buy-in is to present it as an inevitable and necessary part of future infrastructure, and in turn to encourage the adaptation or building of new, anticipatory infrastructures around it."} Widder and Hicks \cite{widderWatchingGenerativeAI2024}
\end{quote}

These radical and synchronous interface and interactive experience modifications are repercussions of an \textit{"AI race for technological advantage"} \cite{caveAIRaceStrategic2018}: ": AI is developed with the intention to reach technological superiority and gain market shares.
Gen-AI services are particularly capital-intensive, which gives the edge to actors that already dominate the digital market. This is tied to an economic approach of rapid product scaling and market capture. If demand is weak or non-existent, it must be stimulated. In this context, the importance of user adoption crystallizes around AI features, which makes the highlighting of AI features in interfaces all the more important. Because AI adoption has been anticipated with massive investments and growth announcements \footnote{https://www.sequoiacap.com/article/ais-600b-question/},\footnote{https://www.ft.com/content/634b7ec5-10c3-44d3-ae49-2a5b9ad566fa}, the use of deceptive patterns is not surprising, as user engagement on AI features is highly strategic. Users are expected to pay, either directly through subscriptions or indirectly through data, the price of the investments in AI systems, and this could remain unnoticed as everything is made to sweep away users’ doubts, disrupt existing digital habits and discourage alternative digital practices.
In this situation, companies might have recourse to deceptive patterns to ensure that users will stay on the service despite a price increase.

In such competitive circumstances, more regulations on fair competition in the digital market could also converge with environmental interests. 
Monopolistic digital platforms are already known for deliberately increasing users' dependency through various means (e.g. lack of interoperability \cite{doctorowCompetitiveCompatibilityLets2021a}, path dependency \cite{arthurCompetingTechnologiesIncreasing1989}, cognitive dependency \cite{bombaertsAttentionEconomyEcology2024}).
This makes it hard for users to leave digital platforms when, for example, they cannot export their data in a format based on an open standard, or when they cannot collaborate or communicate with other people who do not agree to switch platforms. The situation with AI adoption is even more worrying, as AI features are used as an excuse for increasing subscription fees in a saturated market with weak growth perspectives. A good example of this trend is Microsoft 365’s January 2025 announcement of a 30\% fee increase on family subscriptions and 43\% on personal subscriptions~ \footnote{https://www.microsoft.com/en-us/microsoft-365/blog/2025/01/16/copilot-is-now-included-in-microsoft-365-personal-and-family/}. 
As companies continue to invest huge sums in AI deployment~ \footnote{https://www.ft.com/content/634b7ec5-10c3-44d3-ae49-2a5b9ad566fa}, we can reasonably anticipate further price hikes. If users knew in advance that their subscription price could be doubled, would they stay on the service and continue creating files, adding contacts, planning meetings on the service?

In this regard, there is an interesting connection between environmental interests and installed digital market regulations. Legal dispositions such as the Digital Markets Act \footnote{https://digital-markets-act.ec.europa.eu/index\_en} could integrate deceptive patterns to ensure that platforms in the most advantageous market positions are not using deceitful strategies to limit user exodus and taking advantage of lock-ins to dramatically increase subscription prices.

Toczé et al.'s \textit{"lock-ins"} \cite{toczeDarkSideCloud2022} are used in a capitalistic logic that goes against sustainable practices, e.g. by undermining the consumers power to threaten to leave the platform. 
While environmental sustainability might not be the main motivation behind further regulations of market competition in the digital sector, a better regulation of the enforcement of user adoption could give more sustainable alternative platforms a chance.

\subsection{Discouraging moderation in the use of digital services}
Deceptive design patterns can be drawn upon to demonstrate that design choices discourage the adoption or continuation of more sustainable digital uses. Because most products on the digital market are service-based (SAAS), consumers have little power to refuse incremental evolution of the services: cloud-hosted digital services allow companies to transform interfaces overnight and instantly affect all their users. Users who choose to use a platform for certain features, interface quality and/or prices cannot be ensured that all of these characteristics will remain unchanged the next day. In particular, the first strategy we presented describes how interfaces’ affordances and information architecture have evolved and now demand more efforts to pursue digital activities without the use of generative AI: the interfaces’  \textit{choice architecture} \cite{mathurWhatMakesDark2021a} pushes users towards the least sustainable features of the interface, thus degrading the user experience for users who seek to avoid increasing their individual footprint.

For example, the authors of this paper are at the time of writing collaborating on Overleaf and we cautiously dismissed the language suggestions that started to pop up a few months ago. How can we trust that we will still be able to disable such suggestions in the future? If we switch to different software for sustainability reasons, how can we be sure that it will not follow the same path? So far, users have very little control over their software on this matter. In this forced AI transition, tech companies have largely taken advantage of this grey area regarding consumer choice.

We argue that users who chose a software for its features (or because it doesn’t have certain features) should be better protected in order to promote the adoption and continued practice of \textit{moderate digital uses} \cite{widdicksEscapingUnsustainableDigital2022a}, a term that describes digital practices that consume less data in the current growth-oriented paradigm of digital consumption. As deceptive patterns are useful to demonstrate how users’ choices are influenced for corporate purposes, we believe applying the concept to sustainability issues could also be fruitful, by helping to establish whether design choices intentionally or unintentionally support sustainable engagement with the digital service. While placing the responsibility of feature adoption on the user won’t solve the situation in itself, designers and tech companies could be held responsible for ensuring at each update that users could still use the service without increasing their environmental footprint, with no impact on user experience.

\subsection{Greenwashing and lack of transparency}
The EU’s AI Act, which came into effect in August 2024, requires companies to report on the environmental impact of their activities. Sustainability and environmental responsibility reports are actually not new, and are already published by large tech companies every year \footnote{See for example: https://citizen.snap.com/?lang=en-US}. 
While the reliability of self-reported data is questionable and significant limitations exist in the reporting of environmental effects (e.g. disclosing energy consumption or water consumption is not required from the AI Act) \cite{alderAIClimateTransparency2024}, this is a step towards more transparency. Similar arguments  have been put forward  in favour of transparent energy ratings for AI chatbots as a way to encourage minimizing the environmental costs of these services \cite{luccioniLightBulbsHave2024}.

To prevent greenwashing, such as when companies  like Meta~\footnote{\url{https://sustainability.atmeta.com/}} or Microsoft~\footnote{\url{https://cdn-dynmedia-1.microsoft.com/is/content/microsoftcorp/microsoft/msc/documents/presentations/CSR/Microsoft-2024-Environmental-Sustainability-Report.pdf\#page=10}} purport to strive for net-zero emissions and water consumption while actually employing design strategies that maximize such emissions, companies could be held accountable for the way their platform designs steer demand on features toward higher environmental impacts. 
Similarly, energy rating labels do not prevent users from intensifying their AI use. In this sense, we agree with the argument articulated by Kender et al. \cite{kenderShapeSocialMedia2022}, who in the case of social media found that \textit{``while functional design (e.g. algorithmic content suggestion, data collection) is increasingly being regulated, aesthetic design is barely considered in discussions about the ef[f]ects''}. What they call \textit{``aesthetic design power''} could be harnessed to address situations in which design is employed to offset  or contradict ethical or legal corporate commitments.

\section{Future work}
This work aimed at documenting a special moment of the longer transformations effected through interface and interaction design in relation to AI use. The applications and software we studied are changing so rapidly that part of our corpus has already been obsolete as we submit are submitting this article. Further work could explore how AI-based features evolve and are updated, and how strategies and deceptive patterns evolve or develop in relation to generative AI diffusion.
We also purposely didn’t focus on companies or designers’ intents in producing deceptive patterns. The design of AI-based features in these forms could be driven by different motives, including an effort to make AI technologies accessible for a new public. However, it is likely that the context of surveillance capitalism in which these AI-based features are developed highly influences their design, as this economic model relies on mass data collection, not only to offer custom features to users, but also to sell better-profiled data to advertisers. Further studies could uncover the different motivations that explain why AI-based features are designed the way they are.
Future work could also investigate users' reception of the strategies we identified: Did they notice them? How do these interface choices impact the ways users engage in digital systems?

Finally, we see great potential in further exploring deceptive patterns as a means to bring about debates and regulations on the effects of design choices. Digital design needs to be better regulated to mitigate ICT’s environmental effects, along with other effects, including social and psychological ones.

Aside from impacts at the regulatory level, further research aiming at better characterizing deceptive design patterns regarding sustainability could prove useful to designers looking to avoid perpetuating the growth paradigm of the digital sector.


%% file: tex/6_conclusion.tex
\section{Conclusion}
This work has explored the role of interface and interaction design in the strategic diffusion of AI-based features on prominent digital platforms. We have identified two main strategies that reflect various deceptive design patterns: imposing AI features at the expense of existing non-AI features and promoting misleading technology narratives. In different ways, these strategies steer users towards the adoption of AI-based features by leveraging companies’ dominant position with users and in the digital market. The paper discusses the opportunities offered by the “deceptive patterns” concept for advancing the environmental regulation of the design of digital services.


\section{Acknowledgments}
We thank the reviewers and the attendees of the LIMITS conference for their appreciation and constructive comments. We are also grateful for Jean-Yves Bart's careful proofreading and language corrections. Special gratitude to the many people who shared with us AI features. This work was supported by two French government grants managed by the Agence Nationale de la Recherche. References : France 2030 program - ANR-22-EXEN-0006 (PEPR eNSEMBLE / PC5); and SuffisanceNumérique - ANR-23-SSAI-0022 - (https://anr.fr/Projet-ANR-23-SSAI-0022)

%% file: main.bbl

\begin{thebibliography}{46}


\ifx \showCODEN    \undefined \def \showCODEN     #1{\unskip}     \fi
\ifx \showDOI      \undefined \def \showDOI       #1{#1}\fi
\ifx \showISBNx    \undefined \def \showISBNx     #1{\unskip}     \fi
\ifx \showISBNxiii \undefined \def \showISBNxiii  #1{\unskip}     \fi
\ifx \showISSN     \undefined \def \showISSN      #1{\unskip}     \fi
\ifx \showLCCN     \undefined \def \showLCCN      #1{\unskip}     \fi
\ifx \shownote     \undefined \def \shownote      #1{#1}          \fi
\ifx \showarticletitle \undefined \def \showarticletitle #1{#1}   \fi
\ifx \showURL      \undefined \def \showURL       {\relax}        \fi
\providecommand\bibfield[2]{#2}
\providecommand\bibinfo[2]{#2}
\providecommand\natexlab[1]{#1}
\providecommand\showeprint[2][]{arXiv:#2}

\bibitem[Ene(2025)]%
        {EnergyAIAnalysis2025}
 \bibinfo{year}{2025}\natexlab{}.
\newblock \bibinfo{title}{Energy and {{AI}} -- {{Analysis}}}.
\newblock \bibinfo{howpublished}{https://www.iea.org/reports/energy-and-ai}.
\newblock


\bibitem[Alder et~al\mbox{.}(2024)]%
        {alderAIClimateTransparency2024}
\bibfield{author}{\bibinfo{person}{Nicolas Alder}, \bibinfo{person}{Kai Ebert},
  \bibinfo{person}{Ralf Herbrich}, {and} \bibinfo{person}{Philipp Hacker}.}
  \bibinfo{year}{2024}\natexlab{}.
\newblock \bibinfo{title}{{{AI}}, {{Climate}}, and {{Transparency}}:
  {{Operationalizing}} and {{Improving}} the {{AI Act}}}.
\newblock
\newblock
\urldef\tempurl%
\url{https://doi.org/10.48550/arXiv.2409.07471}
\showDOI{\tempurl}
\showeprint[arxiv]{2409.07471}~[cs]


\bibitem[{Apple Inc.}(1987)]%
        {appleinc.EDUCOM1987John1987}
\bibfield{author}{\bibinfo{person}{{Apple Inc.}}}
  \bibinfo{year}{1987}\natexlab{}.
\newblock \bibinfo{title}{{{EDUCOM}} 1987 {{John Sculley Keynote}}}.
\newblock
\newblock


\bibitem[Arthur(1989)]%
        {arthurCompetingTechnologiesIncreasing1989}
\bibfield{author}{\bibinfo{person}{W.~Brian Arthur}.}
  \bibinfo{year}{1989}\natexlab{}.
\newblock \showarticletitle{Competing {{Technologies}}, {{Increasing Returns}},
  and {{Lock-In}} by {{Historical Events}}}.
\newblock \bibinfo{journal}{\emph{The Economic Journal}} \bibinfo{volume}{99},
  \bibinfo{number}{394} (\bibinfo{year}{1989}), \bibinfo{pages}{116--131}.
\newblock
\showISSN{0013-0133}
\urldef\tempurl%
\url{https://doi.org/10.2307/2234208}
\showDOI{\tempurl}
\showeprint[jstor]{2234208}


\bibitem[Baldassarre et~al\mbox{.}(2023)]%
        {baldassarreSocialImpactGenerative2023}
\bibfield{author}{\bibinfo{person}{Maria~Teresa Baldassarre},
  \bibinfo{person}{Danilo Caivano}, \bibinfo{person}{Berenice Fernandez~Nieto},
  \bibinfo{person}{Domenico Gigante}, {and} \bibinfo{person}{Azzurra Ragone}.}
  \bibinfo{year}{2023}\natexlab{}.
\newblock \showarticletitle{The {{Social Impact}} of {{Generative AI}}: {{An
  Analysis}} on {{ChatGPT}}}. In \bibinfo{booktitle}{\emph{Proceedings of the
  2023 {{ACM Conference}} on {{Information Technology}} for {{Social Good}}}}
  \emph{(\bibinfo{series}{{{GoodIT}} '23})}. \bibinfo{publisher}{Association
  for Computing Machinery}, \bibinfo{address}{New York, NY, USA},
  \bibinfo{pages}{363--373}.
\newblock
\showISBNx{979-8-4007-0116-0}
\urldef\tempurl%
\url{https://doi.org/10.1145/3582515.3609555}
\showDOI{\tempurl}


\bibitem[Bennett(2016)]%
        {bennettEnchantmentModernLife2016}
\bibfield{author}{\bibinfo{person}{Jane Bennett}.}
  \bibinfo{year}{2016}\natexlab{}.
\newblock \bibinfo{booktitle}{\emph{The {{Enchantment}} of {{Modern Life}}:
  {{Attachments}}, {{Crossings}}, and {{Ethics}}}}.
\newblock \bibinfo{publisher}{Princeton University Press}.
\newblock
\showISBNx{978-1-4008-8453-7}
\urldef\tempurl%
\url{https://doi.org/10.1515/9781400884537}
\showDOI{\tempurl}


\bibitem[Berthelot et~al\mbox{.}(2024)]%
        {berthelotEstimatingEnvironmentalImpact2024}
\bibfield{author}{\bibinfo{person}{Adrien Berthelot}, \bibinfo{person}{Eddy
  Caron}, \bibinfo{person}{Mathilde Jay}, {and} \bibinfo{person}{Laurent
  Lef{\`e}vre}.} \bibinfo{year}{2024}\natexlab{}.
\newblock \showarticletitle{Estimating the Environmental Impact of
  {{Generative-AI}} Services Using an {{LCA-based}} Methodology}.
\newblock \bibinfo{journal}{\emph{Procedia CIRP}}  \bibinfo{volume}{122}
  (\bibinfo{date}{Jan.} \bibinfo{year}{2024}), \bibinfo{pages}{707--712}.
\newblock
\showISSN{2212-8271}
\urldef\tempurl%
\url{https://doi.org/10.1016/j.procir.2024.01.098}
\showDOI{\tempurl}


\bibitem[Bombaerts et~al\mbox{.}(2024)]%
        {bombaertsAttentionEconomyEcology2024}
\bibfield{author}{\bibinfo{person}{Gunter Bombaerts}, \bibinfo{person}{Tom
  Hannes}, \bibinfo{person}{Martin Adam}, \bibinfo{person}{Alessandra Aloisi},
  \bibinfo{person}{Joel Anderson}, \bibinfo{person}{Lawrence Berger},
  \bibinfo{person}{Stefano~Davide Bettera}, \bibinfo{person}{Enrico Campo},
  \bibinfo{person}{Laura Candiotto}, \bibinfo{person}{Silvia~Caprioglio
  Panizza}, \bibinfo{person}{Yves Citton}, \bibinfo{person}{Diego D.Angelo},
  \bibinfo{person}{Matthew Dennis}, \bibinfo{person}{Nathalie Depraz},
  \bibinfo{person}{Peter Doran}, \bibinfo{person}{Wolfgang Drechsler},
  \bibinfo{person}{Bill Duane}, \bibinfo{person}{William Edelglass},
  \bibinfo{person}{Iris Eisenberger}, \bibinfo{person}{Beverley~Foulks
  McGuire}, \bibinfo{person}{Antony Fredriksson}, \bibinfo{person}{Karamjit~S.
  Gill}, \bibinfo{person}{Peter~D. Hershock}, \bibinfo{person}{Soraj
  Hongladarom}, \bibinfo{person}{Beth Jacobs}, \bibinfo{person}{G{\'a}bor
  Karsai}, \bibinfo{person}{Thomas Lennerfors}, \bibinfo{person}{Jeanne Lim},
  \bibinfo{person}{Chien-Te Lin}, \bibinfo{person}{Mark Losoncz},
  \bibinfo{person}{David Loy}, \bibinfo{person}{Lavinia Marin},
  \bibinfo{person}{Bence~P{\'e}ter Maros{\'a}n}, \bibinfo{person}{Chiara
  Mascarello}, \bibinfo{person}{David McMahan}, \bibinfo{person}{Jin~Y. Park},
  \bibinfo{person}{Nina Petek}, \bibinfo{person}{Anna Puzio},
  \bibinfo{person}{Katrien Schaubroek}, \bibinfo{person}{Jens Schlieter},
  \bibinfo{person}{Brian Schroeder}, \bibinfo{person}{Shobhit Shakya},
  \bibinfo{person}{Juewei Shi}, \bibinfo{person}{Elizaveta Solomonova},
  \bibinfo{person}{Francesco Tormen}, \bibinfo{person}{Jitendra Uttam},
  \bibinfo{person}{Marieke~Van Vugt}, \bibinfo{person}{Sebastjan
  V{\"o}r{\"o}s}, \bibinfo{person}{Maren Wehrle}, \bibinfo{person}{Galit
  Wellner}, \bibinfo{person}{Jason~M. Wirth}, \bibinfo{person}{Olaf Witkowski},
  \bibinfo{person}{Apiradee Wongkitrungrueng}, \bibinfo{person}{Dale~S.
  Wright}, {and} \bibinfo{person}{Yutong Zheng}.}
  \bibinfo{year}{2024}\natexlab{}.
\newblock \bibinfo{title}{From an Attention Economy to an Ecology of Attending.
  {{A}} Manifesto}.
\newblock
\newblock
\urldef\tempurl%
\url{https://doi.org/10.48550/arXiv.2410.17421}
\showDOI{\tempurl}
\showeprint[arxiv]{2410.17421}~[cs]


\bibitem[Borning et~al\mbox{.}(2018)]%
        {borningWhatPushesBack2018}
\bibfield{author}{\bibinfo{person}{Alan Borning}, \bibinfo{person}{Batya
  Friedman}, {and} \bibinfo{person}{Deric Gruen}.}
  \bibinfo{year}{2018}\natexlab{}.
\newblock \showarticletitle{What Pushes Back from Considering Materiality in
  {{IT}}?}. In \bibinfo{booktitle}{\emph{Proceedings of the 2018 {{Workshop}}
  on {{Computing}} within {{Limits}}}} \emph{(\bibinfo{series}{{{LIMITS}}
  '18})}. \bibinfo{publisher}{Association for Computing Machinery},
  \bibinfo{address}{New York, NY, USA}, \bibinfo{pages}{1--6}.
\newblock
\showISBNx{978-1-4503-6575-8}
\urldef\tempurl%
\url{https://doi.org/10.1145/3232617.3232627}
\showDOI{\tempurl}


\bibitem[Borning et~al\mbox{.}(2020)]%
        {borningInvisibleMaterialityInformation2020}
\bibfield{author}{\bibinfo{person}{Alan Borning}, \bibinfo{person}{Batya
  Friedman}, {and} \bibinfo{person}{Nick Logler}.}
  \bibinfo{year}{2020}\natexlab{}.
\newblock \showarticletitle{The 'invisible' Materiality of Information
  Technology}.
\newblock \bibinfo{journal}{\emph{Commun. ACM}} \bibinfo{volume}{63},
  \bibinfo{number}{6} (\bibinfo{date}{May} \bibinfo{year}{2020}),
  \bibinfo{pages}{57--64}.
\newblock
\showISSN{0001-0782}
\urldef\tempurl%
\url{https://doi.org/10.1145/3360647}
\showDOI{\tempurl}


\bibitem[Caporael(1986)]%
        {caporaelAnthropomorphismMechanomorphismTwo1986}
\bibfield{author}{\bibinfo{person}{L.~R. Caporael}.}
  \bibinfo{year}{1986}\natexlab{}.
\newblock \showarticletitle{Anthropomorphism and Mechanomorphism: {{Two}} Faces
  of the Human Machine}.
\newblock \bibinfo{journal}{\emph{Computers in Human Behavior}}
  \bibinfo{volume}{2}, \bibinfo{number}{3} (\bibinfo{date}{Jan.}
  \bibinfo{year}{1986}), \bibinfo{pages}{215--234}.
\newblock
\showISSN{0747-5632}
\urldef\tempurl%
\url{https://doi.org/10.1016/0747-5632(86)90004-X}
\showDOI{\tempurl}


\bibitem[Castelo et~al\mbox{.}(2019)]%
        {casteloTaskDependentAlgorithmAversion2019}
\bibfield{author}{\bibinfo{person}{Noah Castelo}, \bibinfo{person}{Maarten~W.
  Bos}, {and} \bibinfo{person}{Donald~R. Lehmann}.}
  \bibinfo{year}{2019}\natexlab{}.
\newblock \showarticletitle{Task-{{Dependent Algorithm Aversion}}}.
\newblock \bibinfo{journal}{\emph{Journal of Marketing Research}}
  \bibinfo{volume}{56}, \bibinfo{number}{5} (\bibinfo{year}{2019}),
  \bibinfo{pages}{809--825}.
\newblock
\showISSN{0022-2437}
\showeprint[jstor]{26967271}


\bibitem[Cave and {\'O}h{\'E}igeartaigh(2018)]%
        {caveAIRaceStrategic2018}
\bibfield{author}{\bibinfo{person}{Stephen Cave} {and}
  \bibinfo{person}{Se{\'a}n~S. {\'O}h{\'E}igeartaigh}.}
  \bibinfo{year}{2018}\natexlab{}.
\newblock \showarticletitle{An {{AI Race}} for {{Strategic Advantage}}:
  {{Rhetoric}} and {{Risks}}}. In \bibinfo{booktitle}{\emph{Proceedings of the
  2018 {{AAAI}}/{{ACM Conference}} on {{AI}}, {{Ethics}}, and {{Society}}}}
  \emph{(\bibinfo{series}{{{AIES}} '18})}. \bibinfo{publisher}{Association for
  Computing Machinery}, \bibinfo{address}{New York, NY, USA},
  \bibinfo{pages}{36--40}.
\newblock
\showISBNx{978-1-4503-6012-8}
\urldef\tempurl%
\url{https://doi.org/10.1145/3278721.3278780}
\showDOI{\tempurl}


\bibitem[Chordia et~al\mbox{.}(2023)]%
        {chordiaDeceptiveDesignPatterns2023}
\bibfield{author}{\bibinfo{person}{Ishita Chordia},
  \bibinfo{person}{Lena-Phuong Tran}, \bibinfo{person}{Tala~June Tayebi},
  \bibinfo{person}{Emily Parrish}, \bibinfo{person}{Sheena Erete},
  \bibinfo{person}{Jason Yip}, {and} \bibinfo{person}{Alexis Hiniker}.}
  \bibinfo{year}{2023}\natexlab{}.
\newblock \showarticletitle{Deceptive {{Design Patterns}} in {{Safety
  Technologies}}: {{A Case Study}} of the {{Citizen App}}}. In
  \bibinfo{booktitle}{\emph{Proceedings of the 2023 {{CHI Conference}} on
  {{Human Factors}} in {{Computing Systems}}}} \emph{(\bibinfo{series}{{{CHI}}
  '23})}. \bibinfo{publisher}{Association for Computing Machinery},
  \bibinfo{address}{New York, NY, USA}, \bibinfo{pages}{1--18}.
\newblock
\showISBNx{978-1-4503-9421-5}
\urldef\tempurl%
\url{https://doi.org/10.1145/3544548.3581258}
\showDOI{\tempurl}


\bibitem[Chu(2024)]%
        {chuUSSlowsPlans2024}
\bibfield{author}{\bibinfo{person}{Amanda Chu}.}
  \bibinfo{year}{2024}\natexlab{}.
\newblock \showarticletitle{{{US}} Slows Plans to Retire Coal-Fired Plants as
  Power Demand from {{AI}} Surges}.
\newblock \bibinfo{journal}{\emph{Financial Times}} (\bibinfo{date}{May}
  \bibinfo{year}{2024}).
\newblock


\bibitem[Di~Geronimo et~al\mbox{.}(2020)]%
        {digeronimoUIDarkPatterns2020}
\bibfield{author}{\bibinfo{person}{Linda Di~Geronimo}, \bibinfo{person}{Larissa
  Braz}, \bibinfo{person}{Enrico Fregnan}, \bibinfo{person}{Fabio Palomba},
  {and} \bibinfo{person}{Alberto Bacchelli}.} \bibinfo{year}{2020}\natexlab{}.
\newblock \showarticletitle{{{UI Dark Patterns}} and {{Where}} to {{Find
  Them}}}.
\newblock \bibinfo{journal}{\emph{Proceedings of the 2020 CHI Conference on
  Human Factors in Computing Systems}} (\bibinfo{year}{2020}).
\newblock
\urldef\tempurl%
\url{https://doi.org/10.1145/3313831.3376600}
\showDOI{\tempurl}


\bibitem[Doctorow(2021)]%
        {doctorowCompetitiveCompatibilityLets2021a}
\bibfield{author}{\bibinfo{person}{Cory Doctorow}.}
  \bibinfo{year}{2021}\natexlab{}.
\newblock \showarticletitle{Competitive Compatibility: Let's Fix the Internet,
  Not the Tech Giants}.
\newblock \bibinfo{journal}{\emph{Commun. ACM}} \bibinfo{volume}{64},
  \bibinfo{number}{10} (\bibinfo{date}{Oct.} \bibinfo{year}{2021}),
  \bibinfo{pages}{26--29}.
\newblock
\showISSN{0001-0782, 1557-7317}
\urldef\tempurl%
\url{https://doi.org/10.1145/3446789}
\showDOI{\tempurl}


\bibitem[Falk et~al\mbox{.}(2024)]%
        {falkAttributionProblemSeemingly2024}
\bibfield{author}{\bibinfo{person}{Sophia Falk}, \bibinfo{person}{Aimee {van
  Wynsberghe}}, {and} \bibinfo{person}{Lisa {Biber-Freudenberger}}.}
  \bibinfo{year}{2024}\natexlab{}.
\newblock \showarticletitle{The Attribution Problem of a Seemingly Intangible
  Industry}.
\newblock \bibinfo{journal}{\emph{Environmental Challenges}}
  \bibinfo{volume}{16} (\bibinfo{date}{Aug.} \bibinfo{year}{2024}),
  \bibinfo{pages}{101003}.
\newblock
\showISSN{2667-0100}
\urldef\tempurl%
\url{https://doi.org/10.1016/j.envc.2024.101003}
\showDOI{\tempurl}


\bibitem[Gray et~al\mbox{.}(2024)]%
        {grayOntologyDarkPatterns2024}
\bibfield{author}{\bibinfo{person}{Colin~M. Gray},
  \bibinfo{person}{Cristiana~Teixeira Santos}, \bibinfo{person}{Nataliia
  Bielova}, {and} \bibinfo{person}{Thomas Mildner}.}
  \bibinfo{year}{2024}\natexlab{}.
\newblock \showarticletitle{An {{Ontology}} of {{Dark Patterns Knowledge}}:
  {{Foundations}}, {{Definitions}}, and a {{Pathway}} for {{Shared
  Knowledge-Building}}}. In \bibinfo{booktitle}{\emph{Proceedings of the 2024
  {{CHI Conference}} on {{Human Factors}} in {{Computing Systems}}}}
  \emph{(\bibinfo{series}{{{CHI}} '24})}. \bibinfo{publisher}{Association for
  Computing Machinery}, \bibinfo{address}{New York, NY, USA},
  \bibinfo{pages}{1--22}.
\newblock
\showISBNx{979-8-4007-0330-0}
\urldef\tempurl%
\url{https://doi.org/10.1145/3613904.3642436}
\showDOI{\tempurl}


\bibitem[Guidi et~al\mbox{.}(2024)]%
        {guidiEnvironmentalBurdenUnited2024}
\bibfield{author}{\bibinfo{person}{Gianluca Guidi}, \bibinfo{person}{Francesca
  Dominici}, \bibinfo{person}{Jonathan Gilmour}, \bibinfo{person}{Kevin
  Butler}, \bibinfo{person}{Eric Bell}, \bibinfo{person}{Scott Delaney}, {and}
  \bibinfo{person}{Falco~J. {Bargagli-Stoffi}}.}
  \bibinfo{year}{2024}\natexlab{}.
\newblock \bibinfo{title}{Environmental {{Burden}} of {{United States Data
  Centers}} in the {{Artificial Intelligence Era}}}.
\newblock
\newblock
\urldef\tempurl%
\url{https://doi.org/10.48550/arXiv.2411.09786}
\showDOI{\tempurl}
\showeprint[arxiv]{2411.09786}~[cs]


\bibitem[Ibrahim et~al\mbox{.}(2024)]%
        {ibrahimCharacterizingModelingHarms2024}
\bibfield{author}{\bibinfo{person}{Lujain Ibrahim}, \bibinfo{person}{Luc
  Rocher}, {and} \bibinfo{person}{Ana Valdivia}.}
  \bibinfo{year}{2024}\natexlab{}.
\newblock \bibinfo{title}{Characterizing and Modeling Harms from Interactions
  with Design Patterns in {{AI}} Interfaces}.
\newblock
\newblock
\showeprint[arxiv]{2404.11370}~[cs]


\bibitem[Janlert and Stolterman(2017)]%
        {janlertThingsThatKeep2017}
\bibfield{author}{\bibinfo{person}{Lars-Erik Janlert} {and}
  \bibinfo{person}{Erik Stolterman}.} \bibinfo{year}{2017}\natexlab{}.
\newblock \bibinfo{booktitle}{\emph{Things {{That Keep Us Busy}}: {{The
  Elements}} of {{Interaction}}}}.
\newblock \bibinfo{publisher}{The MIT Press}.
\newblock
\showISBNx{978-0-262-03664-1}


\bibitem[Kelly(2018)]%
        {kellyWorksMagicMetaphor2018}
\bibfield{author}{\bibinfo{person}{Nicholas~M. Kelly}.}
  \bibinfo{year}{2018}\natexlab{}.
\newblock \showarticletitle{``{{Works}} like {{Magic}}'': {{Metaphor}},
  {{Meaning}}, and the {{GUI}} in {{Snow Crash}}}.
\newblock \bibinfo{journal}{\emph{Science Fiction Studies}}
  \bibinfo{volume}{45}, \bibinfo{number}{Part 1 (134)} (\bibinfo{date}{March}
  \bibinfo{year}{2018}), \bibinfo{pages}{69--90}.
\newblock
\showISSN{0091-7729}
\urldef\tempurl%
\url{https://doi.org/10.5621/sciefictstud.45.1.0069}
\showDOI{\tempurl}


\bibitem[Kender and Frauenberger(2022)]%
        {kenderShapeSocialMedia2022}
\bibfield{author}{\bibinfo{person}{Kay Kender} {and}
  \bibinfo{person}{Christopher Frauenberger}.} \bibinfo{year}{2022}\natexlab{}.
\newblock \showarticletitle{The {{Shape}} of {{Social Media}}: {{Towards
  Addressing}} ({{Aesthetic}}) {{Design Power}}}. In
  \bibinfo{booktitle}{\emph{Proceedings of the 2022 {{ACM Designing Interactive
  Systems Conference}}}} \emph{(\bibinfo{series}{{{DIS}} '22})}.
  \bibinfo{publisher}{Association for Computing Machinery},
  \bibinfo{address}{New York, NY, USA}, \bibinfo{pages}{365--376}.
\newblock
\showISBNx{978-1-4503-9358-4}
\urldef\tempurl%
\url{https://doi.org/10.1145/3532106.3533470}
\showDOI{\tempurl}


\bibitem[Li et~al\mbox{.}(2025)]%
        {liMakingAILess2025}
\bibfield{author}{\bibinfo{person}{Pengfei Li}, \bibinfo{person}{Jianyi Yang},
  \bibinfo{person}{Mohammad~A. Islam}, {and} \bibinfo{person}{Shaolei Ren}.}
  \bibinfo{year}{2025}\natexlab{}.
\newblock \bibinfo{title}{Making {{AI Less}} "{{Thirsty}}": {{Uncovering}} and
  {{Addressing}} the {{Secret Water Footprint}} of {{AI Models}}}.
\newblock
\newblock
\urldef\tempurl%
\url{https://doi.org/10.48550/arXiv.2304.03271}
\showDOI{\tempurl}
\showeprint[arxiv]{2304.03271}~[cs]


\bibitem[Ligozat et~al\mbox{.}(2022)]%
        {ligozatUnravelingHiddenEnvironmental2022}
\bibfield{author}{\bibinfo{person}{Anne-Laure Ligozat}, \bibinfo{person}{Julien
  Lefevre}, \bibinfo{person}{Aur{\'e}lie Bugeau}, {and}
  \bibinfo{person}{Jacques Combaz}.} \bibinfo{year}{2022}\natexlab{}.
\newblock \showarticletitle{Unraveling the {{Hidden Environmental Impacts}} of
  {{AI Solutions}} for {{Environment Life Cycle Assessment}} of {{AI
  Solutions}}}.
\newblock \bibinfo{journal}{\emph{Sustainability}} \bibinfo{volume}{14},
  \bibinfo{number}{9} (\bibinfo{date}{Jan.} \bibinfo{year}{2022}),
  \bibinfo{pages}{5172}.
\newblock
\showISSN{2071-1050}
\urldef\tempurl%
\url{https://doi.org/10.3390/su14095172}
\showDOI{\tempurl}


\bibitem[Luccioni et~al\mbox{.}(2024b)]%
        {luccioniPowerHungryProcessing2024}
\bibfield{author}{\bibinfo{person}{Alexandra~Sasha Luccioni},
  \bibinfo{person}{Yacine Jernite}, {and} \bibinfo{person}{Emma Strubell}.}
  \bibinfo{year}{2024}\natexlab{b}.
\newblock \showarticletitle{Power {{Hungry Processing}}: {{Watts Driving}} the
  {{Cost}} of {{AI Deployment}}?}. In \bibinfo{booktitle}{\emph{The 2024 {{ACM
  Conference}} on {{Fairness}}, {{Accountability}}, and {{Transparency}}}}.
  \bibinfo{pages}{85--99}.
\newblock
\urldef\tempurl%
\url{https://doi.org/10.1145/3630106.3658542}
\showDOI{\tempurl}
\showeprint[arxiv]{2311.16863}~[cs]


\bibitem[Luccioni et~al\mbox{.}(2025)]%
        {luccioniEfficiencyGainsRebound2025}
\bibfield{author}{\bibinfo{person}{Alexandra~Sasha Luccioni},
  \bibinfo{person}{Emma Strubell}, {and} \bibinfo{person}{Kate Crawford}.}
  \bibinfo{year}{2025}\natexlab{}.
\newblock \bibinfo{title}{From {{Efficiency Gains}} to {{Rebound Effects}}:
  {{The Problem}} of {{Jevons}}' {{Paradox}} in {{AI}}'s {{Polarized
  Environmental Debate}}}.
\newblock
\newblock
\urldef\tempurl%
\url{https://doi.org/10.48550/arXiv.2501.16548}
\showDOI{\tempurl}
\showeprint[arxiv]{2501.16548}~[cs]


\bibitem[Luccioni et~al\mbox{.}(2024a)]%
        {luccioniLightBulbsHave2024}
\bibfield{author}{\bibinfo{person}{Sasha Luccioni}, \bibinfo{person}{Boris
  Gamazaychikov}, \bibinfo{person}{Sara Hooker}, \bibinfo{person}{R{\'e}gis
  Pierrard}, \bibinfo{person}{Emma Strubell}, \bibinfo{person}{Yacine Jernite},
  {and} \bibinfo{person}{Carole-Jean Wu}.} \bibinfo{year}{2024}\natexlab{a}.
\newblock \showarticletitle{Light Bulbs Have Energy Ratings---so Why Can't
  {{AI}} Chatbots?}
\newblock \bibinfo{journal}{\emph{Nature}} \bibinfo{volume}{632},
  \bibinfo{number}{8026} (\bibinfo{year}{2024}), \bibinfo{pages}{736--738}.
\newblock


\bibitem[Lupetti and {Murray-Rust}(2024)]%
        {lupettiUnmakingAIMagic2024}
\bibfield{author}{\bibinfo{person}{Maria~Luce Lupetti} {and}
  \bibinfo{person}{Dave {Murray-Rust}}.} \bibinfo{year}{2024}\natexlab{}.
\newblock \showarticletitle{({{Un}})Making {{AI Magic}}: {{A Design
  Taxonomy}}}. In \bibinfo{booktitle}{\emph{Proceedings of the 2024 {{CHI
  Conference}} on {{Human Factors}} in {{Computing Systems}}}}
  \emph{(\bibinfo{series}{{{CHI}} '24})}. \bibinfo{publisher}{Association for
  Computing Machinery}, \bibinfo{address}{New York, NY, USA},
  \bibinfo{pages}{1--21}.
\newblock
\showISBNx{979-8-4007-0330-0}
\urldef\tempurl%
\url{https://doi.org/10.1145/3613904.3641954}
\showDOI{\tempurl}


\bibitem[Makridakis(2017)]%
        {makridakisForthcomingArtificialIntelligence2017}
\bibfield{author}{\bibinfo{person}{Spyros Makridakis}.}
  \bibinfo{year}{2017}\natexlab{}.
\newblock \showarticletitle{The Forthcoming {{Artificial Intelligence}}
  ({{AI}}) Revolution: {{Its}} Impact on Society and Firms}.
\newblock \bibinfo{journal}{\emph{Futures}}  \bibinfo{volume}{90}
  (\bibinfo{date}{June} \bibinfo{year}{2017}), \bibinfo{pages}{46--60}.
\newblock
\showISSN{0016-3287}
\urldef\tempurl%
\url{https://doi.org/10.1016/j.futures.2017.03.006}
\showDOI{\tempurl}


\bibitem[Mathur et~al\mbox{.}(2021)]%
        {mathurWhatMakesDark2021a}
\bibfield{author}{\bibinfo{person}{Arunesh Mathur}, \bibinfo{person}{Mihir
  Kshirsagar}, {and} \bibinfo{person}{Jonathan Mayer}.}
  \bibinfo{year}{2021}\natexlab{}.
\newblock \showarticletitle{What {{Makes}} a {{Dark Pattern}}... {{Dark}}?
  {{Design Attributes}}, {{Normative Considerations}}, and {{Measurement
  Methods}}}. In \bibinfo{booktitle}{\emph{Proceedings of the 2021 {{CHI
  Conference}} on {{Human Factors}} in {{Computing Systems}}}}
  \emph{(\bibinfo{series}{{{CHI}} '21})}. \bibinfo{publisher}{Association for
  Computing Machinery}, \bibinfo{address}{New York, NY, USA},
  \bibinfo{pages}{1--18}.
\newblock
\showISBNx{978-1-4503-8096-6}
\urldef\tempurl%
\url{https://doi.org/10.1145/3411764.3445610}
\showDOI{\tempurl}


\bibitem[Morand et~al\mbox{.}(2024)]%
        {morandHowGreenCan2024}
\bibfield{author}{\bibinfo{person}{Cl{\'e}ment Morand},
  \bibinfo{person}{Anne-Laure Ligozat}, {and} \bibinfo{person}{Aur{\'e}lie
  N{\'e}v{\'e}ol}.} \bibinfo{year}{2024}\natexlab{}.
\newblock \bibinfo{title}{How {{Green Can AI Be}}? {{A Study}} of {{Trends}} in
  {{Machine Learning Environmental Impacts}}}.
\newblock
\newblock
\urldef\tempurl%
\url{https://doi.org/10.48550/arXiv.2412.17376}
\showDOI{\tempurl}
\showeprint[arxiv]{2412.17376}~[cs]


\bibitem[Murray(2012)]%
        {murrayInventingMediumPrinciples2012}
\bibfield{author}{\bibinfo{person}{Janet~H. Murray}.}
  \bibinfo{year}{2012}\natexlab{}.
\newblock \bibinfo{booktitle}{\emph{Inventing the {{Medium}}: {{Principles}} of
  {{Interaction Design}} as a {{Cultural Practice}}}}.
\newblock \bibinfo{publisher}{The MIT Press}.
\newblock
\showISBNx{978-0-262-01614-8}
\showeprint[jstor]{j.ctt5hhjgg}


\bibitem[Nagy and Neff(2024)]%
        {nagyConjuringAlgorithmsUnderstanding2024}
\bibfield{author}{\bibinfo{person}{Peter Nagy} {and} \bibinfo{person}{Gina
  Neff}.} \bibinfo{year}{2024}\natexlab{}.
\newblock \showarticletitle{Conjuring Algorithms: {{Understanding}} the Tech
  Industry as Stage Magicians}.
\newblock \bibinfo{journal}{\emph{New Media \& Society}} \bibinfo{volume}{26},
  \bibinfo{number}{9} (\bibinfo{date}{Sept.} \bibinfo{year}{2024}),
  \bibinfo{pages}{4938--4954}.
\newblock
\showISSN{1461-4448}
\urldef\tempurl%
\url{https://doi.org/10.1177/14614448241251789}
\showDOI{\tempurl}


\bibitem[Natale and Ballatore(2020)]%
        {nataleImaginingThinkingMachine2020}
\bibfield{author}{\bibinfo{person}{Simone Natale} {and} \bibinfo{person}{Andrea
  Ballatore}.} \bibinfo{year}{2020}\natexlab{}.
\newblock \showarticletitle{Imagining the Thinking Machine: {{Technological}}
  Myths and the Rise of Artificial Intelligence}.
\newblock \bibinfo{journal}{\emph{Convergence}} \bibinfo{volume}{26},
  \bibinfo{number}{1} (\bibinfo{date}{Feb.} \bibinfo{year}{2020}),
  \bibinfo{pages}{3--18}.
\newblock
\showISSN{1354-8565}
\urldef\tempurl%
\url{https://doi.org/10.1177/1354856517715164}
\showDOI{\tempurl}


\bibitem[Newendorp et~al\mbox{.}(2024)]%
        {newendorpApplesKnowledgeNavigator2024}
\bibfield{author}{\bibinfo{person}{Amanda~K. Newendorp},
  \bibinfo{person}{Mohammadamin Sanaei}, \bibinfo{person}{Arthur~J Perron},
  \bibinfo{person}{Hila Sabouni}, \bibinfo{person}{Nikoo Javadpour},
  \bibinfo{person}{Maddie Sells}, \bibinfo{person}{Katherine Nelson},
  \bibinfo{person}{Michael Dorneich}, {and} \bibinfo{person}{Stephen~B.
  Gilbert}.} \bibinfo{year}{2024}\natexlab{}.
\newblock \showarticletitle{Apple's {{Knowledge Navigator}}: {{Why Doesn}}'t
  That {{Conversational Agent Exist Yet}}?}. In
  \bibinfo{booktitle}{\emph{Proceedings of the 2024 {{CHI Conference}} on
  {{Human Factors}} in {{Computing Systems}}}} \emph{(\bibinfo{series}{{{CHI}}
  '24})}. \bibinfo{publisher}{Association for Computing Machinery},
  \bibinfo{address}{New York, NY, USA}, \bibinfo{pages}{1--14}.
\newblock
\showISBNx{979-8-4007-0330-0}
\urldef\tempurl%
\url{https://doi.org/10.1145/3613904.3642739}
\showDOI{\tempurl}


\bibitem[Pasquinelli et~al\mbox{.}(2024)]%
        {pasquinelliAIWorkAutomation2024}
\bibfield{author}{\bibinfo{person}{Matteo Pasquinelli},
  \bibinfo{person}{Cristina Alaimo}, {and} \bibinfo{person}{Alessandro
  Gandini}.} \bibinfo{year}{2024}\natexlab{}.
\newblock \showarticletitle{{{AI}} at {{Work}}: {{Automation}}, {{Distributed
  Cognition}}, and {{Cultural Embeddedness}}}.
\newblock \bibinfo{journal}{\emph{Tecnoscienza -- Italian Journal of Science
  \&amp; Technology Studies}} \bibinfo{volume}{15}, \bibinfo{number}{1}
  (\bibinfo{date}{July} \bibinfo{year}{2024}), \bibinfo{pages}{99--131}.
\newblock
\urldef\tempurl%
\url{https://doi.org/10.6092/ISSN.2038-3460/20010}
\showDOI{\tempurl}


\bibitem[Preist et~al\mbox{.}(2016)]%
        {preistUnderstandingMitigatingEffects2016}
\bibfield{author}{\bibinfo{person}{Chris Preist}, \bibinfo{person}{Daniel
  Schien}, {and} \bibinfo{person}{Eli Blevis}.}
  \bibinfo{year}{2016}\natexlab{}.
\newblock \showarticletitle{Understanding and {{Mitigating}} the {{Effects}} of
  {{Device}} and {{Cloud Service Design Decisions}} on the {{Environmental
  Footprint}} of {{Digital Infrastructure}}}. In
  \bibinfo{booktitle}{\emph{Proceedings of the 2016 {{CHI Conference}} on
  {{Human Factors}} in {{Computing Systems}}}}. \bibinfo{publisher}{ACM},
  \bibinfo{address}{San Jose California USA}, \bibinfo{pages}{1324--1337}.
\newblock
\showISBNx{978-1-4503-3362-7}
\urldef\tempurl%
\url{https://doi.org/10.1145/2858036.2858378}
\showDOI{\tempurl}


\bibitem[Sasha~Luccioni et~al\mbox{.}(2022)]%
        {sashaluccioniEstimatingCarbonFootprint2022}
\bibfield{author}{\bibinfo{person}{Alexandra Sasha~Luccioni},
  \bibinfo{person}{Sylvain Viguier}, {and} \bibinfo{person}{Anne-Laure
  Ligozat}.} \bibinfo{year}{2022}\natexlab{}.
\newblock \bibinfo{title}{Estimating the {{Carbon Footprint}} of {{BLOOM}}, a
  {{176B Parameter Language Model}}}.
\newblock
\newblock
\urldef\tempurl%
\url{https://doi.org/10.48550/arXiv.2211.02001}
\showDOI{\tempurl}


\bibitem[Strubell et~al\mbox{.}(2020)]%
        {strubellEnergyPolicyConsiderations2020}
\bibfield{author}{\bibinfo{person}{Emma Strubell}, \bibinfo{person}{Ananya
  Ganesh}, {and} \bibinfo{person}{Andrew McCallum}.}
  \bibinfo{year}{2020}\natexlab{}.
\newblock \showarticletitle{Energy and {{Policy Considerations}} for {{Modern
  Deep Learning Research}}}.
\newblock \bibinfo{journal}{\emph{Proceedings of the AAAI Conference on
  Artificial Intelligence}} \bibinfo{volume}{34}, \bibinfo{number}{09}
  (\bibinfo{date}{April} \bibinfo{year}{2020}), \bibinfo{pages}{13693--13696}.
\newblock
\showISSN{2374-3468}
\urldef\tempurl%
\url{https://doi.org/10.1609/aaai.v34i09.7123}
\showDOI{\tempurl}


\bibitem[Tocz{\'e} et~al\mbox{.}(2022)]%
        {toczeDarkSideCloud2022}
\bibfield{author}{\bibinfo{person}{Klervie Tocz{\'e}},
  \bibinfo{person}{Ma{\"e}l Madon}, \bibinfo{person}{Muriel Garcia}, {and}
  \bibinfo{person}{Patricia Lago}.} \bibinfo{year}{2022}\natexlab{}.
\newblock \showarticletitle{The {{Dark Side}} of {{Cloud}} and {{Edge
  Computing}}: {{An Exploratory Study}}}. In
  \bibinfo{booktitle}{\emph{Computing within {{Limits}}}}.
  \bibinfo{publisher}{LIMITS}.
\newblock
\urldef\tempurl%
\url{https://doi.org/10.21428/bf6fb269.9422c084}
\showDOI{\tempurl}


\bibitem[Varoquaux et~al\mbox{.}(2025)]%
        {varoquauxHypeSustainabilityPrice2025}
\bibfield{author}{\bibinfo{person}{Ga{\"e}l Varoquaux},
  \bibinfo{person}{Alexandra~Sasha Luccioni}, {and} \bibinfo{person}{Meredith
  Whittaker}.} \bibinfo{year}{2025}\natexlab{}.
\newblock \bibinfo{title}{Hype, {{Sustainability}}, and the {{Price}} of the
  {{Bigger-is-Better Paradigm}} in {{AI}}}.
\newblock
\newblock
\urldef\tempurl%
\url{https://doi.org/10.48550/arXiv.2409.14160}
\showDOI{\tempurl}
\showeprint[arxiv]{2409.14160}~[cs]


\bibitem[Widder and Hicks(2024)]%
        {widderWatchingGenerativeAI2024}
\bibfield{author}{\bibinfo{person}{David~Gray Widder} {and}
  \bibinfo{person}{Mar Hicks}.} \bibinfo{year}{2024}\natexlab{}.
\newblock \bibinfo{title}{Watching the {{Generative AI Hype Bubble Deflate}}}.
\newblock
\newblock
\urldef\tempurl%
\url{https://doi.org/10.48550/arXiv.2408.08778}
\showDOI{\tempurl}
\showeprint[arxiv]{2408.08778}~[cs]


\bibitem[Widdicks et~al\mbox{.}(2022)]%
        {widdicksEscapingUnsustainableDigital2022a}
\bibfield{author}{\bibinfo{person}{Kelly Widdicks}, \bibinfo{person}{Christian
  Remy}, \bibinfo{person}{Oliver Bates}, \bibinfo{person}{Adrian Friday}, {and}
  \bibinfo{person}{Mike Hazas}.} \bibinfo{year}{2022}\natexlab{}.
\newblock \showarticletitle{Escaping Unsustainable Digital Interactions:
  {{Toward}} ``More Meaningful'' and ``Moderate'' Online Experiences}.
\newblock \bibinfo{journal}{\emph{International Journal of Human-Computer
  Studies}}  \bibinfo{volume}{165} (\bibinfo{date}{Sept.}
  \bibinfo{year}{2022}), \bibinfo{pages}{102853}.
\newblock
\showISSN{10715819}
\urldef\tempurl%
\url{https://doi.org/10.1016/j.ijhcs.2022.102853}
\showDOI{\tempurl}


\bibitem[Zhang and Gosline(2023)]%
        {zhangHumanFavoritismNot2023}
\bibfield{author}{\bibinfo{person}{Yunhao Zhang} {and}
  \bibinfo{person}{Ren{\'e}e Gosline}.} \bibinfo{year}{2023}\natexlab{}.
\newblock \showarticletitle{Human Favoritism, Not {{AI}} Aversion: {{People}}'s
  Perceptions (and Bias) toward Generative {{AI}}, Human Experts, and
  Human--{{GAI}} Collaboration in Persuasive Content Generation}.
\newblock \bibinfo{journal}{\emph{Judgment and Decision Making}}
  \bibinfo{volume}{18} (\bibinfo{year}{2023}), \bibinfo{pages}{e41}.
\newblock
\showISSN{1930-2975}
\urldef\tempurl%
\url{https://doi.org/10.1017/jdm.2023.37}
\showDOI{\tempurl}


\end{thebibliography}
